\documentclass[iop,numberedappendix,appendixfloats,twocolappendix]{emulateapj}

\usepackage{graphicx,epsfig,natbib}
\gdef\msun{${\rm M}_{\odot}$}

\def\Msun{M_\odot/h^2}
\def\Lsun{L_\odot/h^2}

\def\Mi{$M_{i}$}
\def\Mih{$M_{i} - 5logh$}

\gdef\Mstar{M$_{*}$}

\gdef\Re{$R_{e}$}

\slugcomment{}

\shorttitle{MaNGA sample design}
\shortauthors{Wake et al.}

\begin{document}


\title{The SDSS-IV MaNGA Sample: Design, Optimization, and Usage Considerations}


\author{David A. Wake\altaffilmark{1,2,3}, Kevin Bundy\altaffilmark{4,5}, Aleksandar M. Diamond-Stanic\altaffilmark{6,2}, Renbin Yan\altaffilmark{7}, Michael R. Blanton\altaffilmark{8}, Matthew A. Bershady\altaffilmark{2}, Jos\'e R.~S\'anchez-Gallego\altaffilmark{9}, Niv Drory\altaffilmark{10}, Amy Jones\altaffilmark{11}, Guinevere Kauffmann\altaffilmark{11}, David R. Law\altaffilmark{12}, Cheng Li\altaffilmark{13,14}, Nicholas MacDonald\altaffilmark{9}, Karen Masters\altaffilmark{15,16}, Daniel Thomas\altaffilmark{15,16}, Jeremy Tinker\altaffilmark{8}, Anne-Marie Weijmans\altaffilmark{17}, Joel R. Brownstein\altaffilmark{18}}


\altaffiltext{1}{School of Physical Sciences, The Open University, Milton Keynes,  MK7 6AA UK;david.wake@open.ac.uk}
\altaffiltext{2}{Astronomy Department, University of Wisconsin-Madison, Madison, WI 53706, USA}
\altaffiltext{3}{Department of Physics, University of North Carolina Asheville, One University Heights, Asheville, NC 28804, USA}
\altaffiltext{4}{Dept. of Astronomy and Astrophysics, UC Santa Cruz, MS: UCO / LICK, 1156 High St, Santa Cruz, CA 95064, USA}
\altaffiltext{5}{Kavli IPMU (WPI), UTIAS, The University of Tokyo, Kashiwa, Chiba 277-8583, Japan}
\altaffiltext{6}{Department of Physics and Astronomy, Bates College, 44 Campus Avenue, Carnegie Science Hall, Lewiston, Maine 04240, USA}
\altaffiltext{7}{Department of Physics and Astronomy, University of Kentucky, 505 Rose St., Lexington, KY 40506-0057, USA}
\altaffiltext{8}{Center for Cosmology and Particle Physics, Department of Physics, New York University, 4 Washington Place, NY 10003, New York, USA}
\altaffiltext{9}{Department of Astronomy, Box 351580, University of Washington, Seattle, WA 98195, USA}
\altaffiltext{10}{McDonald Observatory, University of Texas at Austin, 1 University Station, Austin, TX 78712-0259, USA}
\altaffiltext{11}{Max-Planck Institut für Astrophysik, D-85741 Garching, Germany}
\altaffiltext{12}{Space Telescope Science Institute, 3700 San Martin Drive, Baltimore, MD 21218, USA}
\altaffiltext{13}{Department of Physics and Tsinghua Center for Astrophysics, Tsinghua University, Beijing 100084, China}
\altaffiltext{14}{Shanghai Astronomical Observatory, Nandan Road 80, Shanghai 200030, China}
\altaffiltext{15}{Institute of Cosmology and Gravitation, University of Portsmouth, Portsmouth, UK}
\altaffiltext{16}{SEPnet, South East Physics Network ({\tt www.sepnet.ac.uk})}
\altaffiltext{17}{School of Physics and Astronomy, University of St Andrews, North Haugh, St Andrews KY16 9SS, UK}
\altaffiltext{18}{Department of Physics and Astronomy, University of Utah, 115 S. 1400 E., Salt Lake City, UT 84112, USA}


\begin{abstract}

We describe the sample design for the SDSS-IV MaNGA survey and present the final properties of the main samples along with important considerations for using these samples for science. Our target selection criteria were developed while simultaneously optimizing the size distribution of the MaNGA integral field units (IFUs), the IFU allocation strategy, and the target density to produce a survey defined in terms of maximizing S/N, spatial resolution, and sample size. Our selection strategy makes use of redshift limits that only depend on $i$-band absolute magnitude (\Mi), or, for a small subset of our sample, \Mi~and color ($NUV-i$).  Such a strategy ensures that all galaxies span the same range in angular size irrespective of luminosity and are therefore covered evenly by the adopted range of IFU sizes. We define three samples: the Primary and Secondary samples are selected to have a flat number density with respect to \Mi~and are targeted to have spectroscopic coverage to 1.5 and 2.5 effective radii (\Re), respectively. The Color-Enhanced supplement increases the number of galaxies in the low-density regions of color-magnitude space by extending the redshift limits of the Primary sample in the appropriate color bins. The samples cover the stellar mass range $5\times10^8 \leq M_* \leq 3\times10^{11} \Msun$ and are sampled at median physical resolutions of 1.37 kpc and 2.5 kpc for the Primary and Secondary samples respectively. We provide weights that will statistically correct for our luminosity and color-dependent selection function and IFU allocation strategy, thus correcting the observed sample to a volume limited sample.

\end{abstract}





\section{Introduction}
\label{sec:intro}

The SDSS-IV MaNGA survey \citep{Bundy15, Blanton17} is using the ARC 2.5m telescope \citep{Gunn06} and
the BOSS spectrographs \citep{Smee13} with its fibers bundled into
multiple IFUs \citep{Drory15} to measure spatially resolved
spectroscopy of $\sim$10,000 nearby galaxies. We have chosen to target
a well defined sample that has uniform spatial coverage in units of
$r$-band effective radius along the major axis ($R_e$), and an
approximately flat stellar mass distribution with $10^9 \lesssim
M_*/\Msun \lesssim 10^{11}$. In this paper, we discuss the motivation
and methodology of the MaNGA sample selection, and we present the
resulting sample in a way that allows for its use in statistical
analysis of galaxy properties.

The challenge of designing a survey like MaNGA is to balance the need
for sample size, spatial coverage, and spatial resolution; these three
parameters compete with each other for finite fiber resources. We have
chosen a sweet spot in this multi-parameter space that best matches
our science requirements \citep[outlined in][]{Bundy15,Yan16} in the
context of a six-year survey duration, existing spectrographs, and
telescope field of view. Since the sample design and the modifications
to the BOSS spectrographs' fiber feeds \citep{Drory15} occurred
concurrently we were able to optimize both together to a considerable
degree. Specifically, we determined the optimal IFU size complement
within the confines of a total fiber budget and viable sample
design. Fortuitously, the redshift range $0.02 \lesssim z \lesssim 0.1$ that balances angular size versus resolution also delivers a target surface density
that is well matched to the telescope field of view (3 degrees in
diameter) and the roughly 1500 fibers with 2\arcsec\ diameters of
MaNGA's feed to the BOSS spectrographs. While we had not foreseen how
well matched the telescope and instrument `grasp' were to our
optimized target density, in hind-sight it is a lesson learned for
planning future surveys. One of the aims of this paper is to
demonstrate how, with adequate knowledge of target density,
well-matched instrumentation can be optimally configured to achieve
well-motivated survey science requirements.

A number of our design choices, such as an even sampling in stellar mass, roughly uniform radial coverage, and a sample size in the thousands, are similar in spirit to those of the SAMI survey \citep{Croom12,Bryant15}. Such choices result naturally from a desire to efficiently study the local galaxy population and produce several similar features in the sample selection approach, such as a stellar mass dependent redshift range. However, our ability to simultaneously design the IFU size distribution and sample selection using a telescope with a larger field does offer further advantages for optimization.

\subsection{Design Strategy}
\label{sec:design}

A number of strategic and tactical choices inform technical elements
of the sample design. A starting point was to select from the well
understood SDSS Main Sample \citep{Strauss02} with enhanced redshift
completeness and remeasured photometry, as described in Section
\ref{sec:cats}. Because the redshifts and global properties of SDSS
galaxies are well known, the distributions of these properties in the
final MaNGA sample can be carefully constructed by effectively
weighting the MaNGA selection in order to maximize its scientific
utility.

\begin{enumerate}
\setcounter{enumi}{0}

\item Sample size: Paramount is the requirement for a large,
  statistically powerful sample size, a choice that comes at the
  expense of higher quality data for individual galaxies within the
  sample. As described in \citet{Bundy15}, the specific argument for
  sampling 10,000 galaxies arises from the desire to divide galaxies
  into $6^3$ groups of $\sim$ 50 galaxies each. These groups, or bins
  (i) sample each of three ``principal components'' defining galaxy
  populations -- stellar mass, SFR and environment; (ii) divide each
  ``dimension'' into 6 bins, sufficient to distinguish the functional
  form of trends across each dimension; and finally (iii) contain
  adequate counting statistics (galaxies) such that differences in
  mean properties between bins can be detected at the 5 sigma level
  even when the measurement precision for individual galaxies is
  comparable to this difference.  This optimization dovetails MaNGA's
  scientific goals for statistical analyses of resolved galaxy
  samples, and complements existing, smaller data sets such as ATLAS3D
  \citep{Cappellari11}, DiskMass \citep{Bershady10}, and CALIFA
  \citep{Sanchez12}, as well as forthcoming data from instruments such
  as MUSE \citep{Bacon10} and KCWI \citep{Martin10} capable of
  producing even higher fidelity data for more modest samples.

\item Sampling in stellar mass: We desire the MaNGA sample to have a
  roughly flat distribution in $\log M_*$ so that studies of
  mass-dependent trends could make use of adequate numbers of
  high-mass galaxies compared to more numerous low-mass systems. A
  flat stellar mass distribution requires an upper redshift limit that
  is stellar mass dependent, so a larger volume is sampled for rarer
  high mass galaxies.

\item Radial coverage: We desire roughly uniform radial coverage as
  defined by some multiple of the effective radius.  This choice is
  motivated by the existence of well-known scaling relations that
  emphasize the importance of the relative length scale of galaxy
  stellar density profiles. Uniform spatial coverage in units of $R_e$
  requires a lower redshift limit that is stellar mass dependent, so
  larger more massive galaxies have the same angular size as smaller
  lower mass galaxies. MaNGA therefore samples the same relative
  extent of the declining surface brightness profile, but at the cost
  of {\it not} maintaining the same {\em physical} spatial resolution
  across the sample.

\item Maximize spatial resolution and S/N: With current facilities, we wish to build a data-set of IFU spectroscopy for
10,000 galaxies with the maximum possible per galaxy physical spatial
resolution, spectral coverage and resolution, and S/N per spatial
element.   
These requirements lead
to several inevitable tactical features of the selection criteria:

\begin{enumerate}

\item to maximize the spatial resolution and total S/N requires the
  selection of galaxies at as low redshift as possible.

\item to reach our goal of 10,000 galaxies requires a sufficiently
  broad redshift distribution so as to have enough galaxies per plate
  to maximize efficiency in IFU allocation.

\end{enumerate}

\end{enumerate}

\subsubsection{Subsamples}

The question of how to set the target radius motivated significant
thought during the sample design.  Smaller multiples of the effective
radius would yield greater spatial resolution and more spatial samples
with higher S/N.  Larger radial coverage would contain more of the
galaxy's light, reach into the dark-matter dominated regime, and probe
unchartered territory in the outskirts of galaxies.  After studying a
number of options, a compromise was reached to cover out to $1.5 R_e$
(the majority of the light distribution) for two-thirds of the sample
and to cover out to $2.5 R_e$ for one-third of the sample. Going to
larger radii, while compelling, was deemed too costly in terms of the
number of spatial samples per IFU with very low S/N.  The sample split
was motivated by basic binning arguments \citep[see][]{Bundy15} and
officially adopted by the science team after the first year of
observations.

With the main sample roughly flat in $\log M_*$, it was possible to
consider a further optimization, that is balancing the rest-frame
color distribution (a proxy for star formation rate) at fixed $M_*$.
In this way, rare populations of star-forming massive galaxies and
non-star-forming low-mass galaxies could be upweighted in the final
sample.  The primary objection was a concern that unexpected biases
could be introduced into the sample and more generally that the
selection would become unnecessary complicated.  As described below, a
practical solution was discovered, however, that helps balance the
color distribution through an additional and modest ``Color-Enhanced
supplement.''  Should it prove biased or undesirable, the supplemental
sample could be easily separated from the Primary sample, and in the
worst case scenario, even ignored.  With the risk mitigated, the
decision was made to include the Color-Enhanced supplement in the
selection.

To summarize, the final full MaNGA sample with which we began the
survey consists of three main subsamples. The Primary sample, which
will initially make up 50\% of the targets, is designed to be covered
by our IFUs to 1.5 \Re~and has a flat distribution in K-corrected
i-band absolute magnitude (\Mi). The Secondary sample, making up 33\%
of the initial targets, is again designed to have a flat distribution
in \Mi~but with coverage to 2.5 \Re. Finally, the Color-Enhanced
supplement is designed to add galaxies in regions of the $NUV-i$ versus
\Mi~color magnitude plane that are under-represented in the Primary
sample, such as high mass blue galaxies and low mass red galaxies, and
will make up 17\% of the initial targets. The combination of the
Primary and Color-Enhanced samples is called the Primary+ sample.

This complexity leads to the final strategic choice in the survey design: 

\begin{enumerate}
\setcounter{enumi}{4}
\item Selection simplicity: 
While we have described the basic strategic and tactical motivations
behind various choices for the sample design, we were also driven to
make the selection as simple and reproducible as possible.  The
implementation of the ``weighting'' described above to deliver a MaNGA
sample with desired global distributions is carried out entirely
through a set of selection criteria involving basic observables that
are relatively model-independent: redshift, $i$-band luminosity, and,
for the Color-Enhanced supplement, (NUV - $i$) color.  Note that the
selection does not depend on effective radius explicitly (although a
radius estimate is used when choosing what sized IFU to allocate to
given galaxy target).  We also emphasize that while much of the sample
design studies made use of $M_*$ estimates, the final selection
employs $i$-band absolute magnitudes as a proxy for
$M_*$.\footnote{For the initial IFU size distribution optimization
  process we used the stellar mass as estimated by the kcorrect code
  \citep{BlantonR07} applied to the five band SDSS photometry. For the final
  samples we have switched to using just i-band absolute magnitude in
  order to simplify the selection function (see Section
  \ref{sec:selfn})} We did not use $M_*$ estimates specifically in
order to avoid potential systematic biases and the use of a
``black-box'' estimator that may be difficult to reproduce.

\end{enumerate}

\subsection{Extant Instrumentation}
\label{sec:instrument}

Various aspects of the sample design are dependent on the nature of
the MaNGA instrumentation.  We highlight a few details here and refer
to \citet{Drory15} for more details.

The MaNGA instrumentation suite is composed of fiber-bundle IFUs
dedicated to observing galaxy targets, with a number of additional
IFUs and single fibers reserved for calibration.  The total number of
fibers, 1423, is limited by the size of the inherited BOSS spectrographs.
The science IFUs contain circular, buffered optical fibers tightly
arranged in a hexagonal format.  This geometry enables IFUs of
different sizes, with specific numbers of fibers for each IFU size.
With a ``live-core'' fiber diameter of 2\arcsec\ and full outer
diameter of 2\farcs5, the smallest science IFUs contains a central
fiber and two outer, hexagonal rings for a total of 19 fibers and
long-axis IFU diameter of 12\farcs5.  Other possible IFU sizes are 37
fibers (17\farcs5), 61 fibers (22\farcs5), 91 fibers (27\farcs5), 127
fibers (32\farcs5), 169 fibers (37\farcs5), 217 fibers (42\farcs5),
and so on.  Choosing the largest IFU size as well as the optimal
distribution of IFU sizes is a major focus of this paper.

Identical sets of the MaNGA instrumentation suite are installed in six
SDSS ``cartridges.''  These sturdy, cylindrical structures house the
light-collecting IFUs and fibers, the field-specific plug plate, and
the output pseudo-slit, which is directed into the spectrographs when the
cartridge is mounted on the telescope. The ferrules and jacketing of
single fibers and MaNGA IFUs are similar, with dimensions that
facilitate hand-plugging of these elements into pre-drilled plates.
As a result there is a ``collision radius'' that defines the minimum
distance between plugged elements.  For the MaNGA IFUs this distance
is 120\arcsec. The mounting of the plate in the cartridge makes use of a post that attaches to the center of the plate helping to deform the plate to the shape of the focal plane. This center post introduces a second ``collision radius'' about the center of the plate of 150\arcsec.

The balance of this paper is organized as follows: In
\S\ref{sec:cats} we describe the construction of the parent catalogs
from the NASA-Sloan Atlas. In \S\ref{sec:zcuts} we describe the
process by which we select the upper and lower redshift cuts for our
Primary and Secondary samples. In \S\ref{sec:bunsizedist}
and \S\ref{sec:maxsize} we describe the methodology that we use to
optimize the IFU size distribution. In \S\ref{sec:density} we describe
how we select the sample space density. In \S\ref{sec:finalsamp} we
describe the results of applying these processes, the selection of the
Color-Enhanced Supplement and detail the properties of the final
samples.  In \S\ref{sec:tiling} we describe how we tile the survey
area and allocate IFUs to the targets. In \S\ref{sec:considerations}
we discuss how to use the sample for statistical analyses of MaNGA
data.

Where applicable we use a flat Lamda-CDM cosmology with $\Omega_M = 0.3$ and $H_0 = 70 km s^{-1} Mpc^{-1}$ except for absolute magnitudes and stellar mass, which are calculated assuming $H_0 = 100 h km s^{-1} Mpc^{-1}$ with $h=1$, following previous versions of the NSA.

\section{Parent Catalogs}
\label{sec:cats}

\begin{figure*}[ht]
 \begin{center}
 \includegraphics[width=12cm,angle=0]{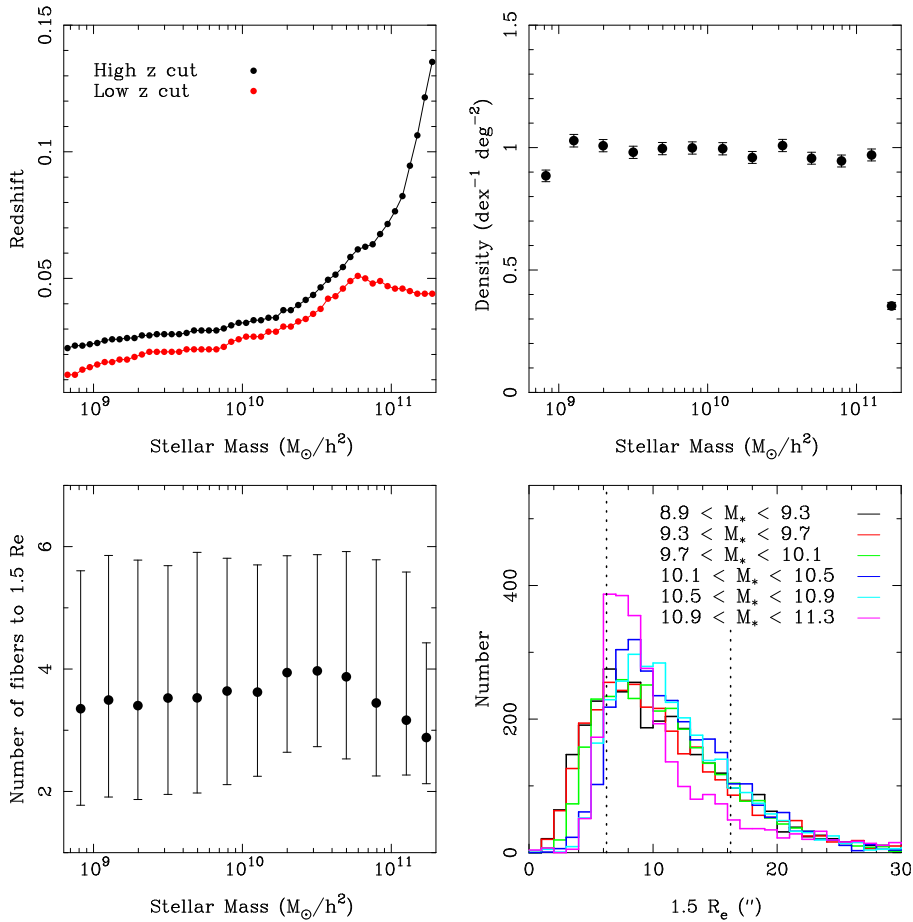}
 \caption{\small A demonstration of the sample selection
   methodology. The top left panel shows the lowest redshift interval
   at each stellar mass that will produce a sample of galaxies where
   80\% can be covered to 1.5 R$_e$ by the 127-fiber IFU (the largest available in this simulation) with a flat number density distribution as a function of 
   stellar mass with a density of 1 deg$^{-2}$ log(M$_*$)$^{-1}$. The
   top right panel shows the resulting stellar mass distribution when
   these cuts are applied to the parent catalog. The bottom left panel
   shows the distribution of the number of 2.5\arcsec-spaced fibers
   that are required to cover 1.5 R$_e$ of each galaxy. The points
   show the median and the error bars show the 20th and 80th
   percentiles. The central fiber is not counted, i.e., the 127-fiber
   IFU has 6 radial fibers. The bottom right panel shows the resulting
   angular size distribution of 1.5 R$_e$ in bins of stellar mass. The
   distributions are very similar regardless of stellar mass, with the
   exception of the most massive galaxies. The vertical dotted lines
   show the radii of the 19- and 127-fiber IFUs.}
  \label{fig:smzcuts}
 \end{center}
 \end{figure*}

The primary input for the selection of all MaNGA galaxies is an enhanced version of the NASA-Sloan Atlas (NSA; Blanton M. http://www.nsatlas.org). The NSA is a catalog of nearby galaxies within 200 Mpc (z $\simeq$ 0.055), primarily based on the SDSS DR7 MAIN galaxy sample \citep{Abazajian09}, but incorporating data from additional sources. The SDSS imaging has been reprocessed to be better suited to the analysis of these large nearby galaxies \citep{Blanton11}. In particular it has improved background subtraction and deblending more suited to nearby large galaxies resulting in more accurate size and luminosity measurements for such galaxies. In addition to a reanalysis of the SDSS imaging, a similar analysis is applied to the GALEX near and far-UV images, and several derived parameters, such as K-corrections and absolute magnitudes\footnote{K-corrections in the NSA catalog do not account for extinction explicitly, and make no attempt to apply an inclination-dependent extinction correction.} (using kcorrect v4\_3), Sersic profile fits, and stellar masses are determined.

The NSA also provides a $\sim 30\%$ improvement in spectroscopic completeness over the standard SDSS spectroscopic catalog for the very brightest sources by adding in redshifts from the NASA Extragalactic Database (NED\footnote{https://ned.ipac.caltech.edu/}), the CfA Redshift Survey (ZCAT\footnote{https://www.cfa.harvard.edu/~dfabricant/huchra/zcat/}, \citealt{Huchra91}), Arecibo Legacy Fast ALFA Survey (ALFALFA, \citealt{Giovanelli05}), the 2dF Galaxy Redshift Survey \citep[2dF][]{Colless01}, and the 6dF Galaxy Redshift Survey \citep[6dF][]{Jones09}. The SDSS is 70\% complete at $r_{\rm AB}\sim14$ and 95\% complete at $r_{\rm AB}\sim16$, emphasizing the importance of these other redshift sources. Assuming the incompleteness of SDSS is orthogonal to the incompleteness of the other sources, we estimate the completeness of the combined sample between $13<r_{\rm AB} <17$ is 98.6\%.

In order to achieve our primary sample design goals (radial coverage and stellar mass range), we need to target massive galaxies at $z >$ 0.055. We have thus extended the NSA analysis to include galaxies with $z <$ 0.15.

We have made one further addition to the standard NSA analysis, the calculation of elliptical Petrosian magnitudes and profiles for all seven bands. The elliptical Petrosian method uses a set of elliptical annuli defined using an estimate of the axis ratio {\it b/a} (minor over major) and the position angle $\phi$. Otherwise it uses the standard algorithm for Petrosian magnitudes, with the Petrosian radius $r_P$ defined as the major axis of the ellipse where the Petrosian ratio $\eta$ = 0.2, and with the aperture for the flux defined with a major-axis radius $2r_P$. $R_e$ is defined as the major axis radius of the ellipse that contains 50\% of the flux within $2r_P$. The NSA pipeline produces several estimates of {\it b/a} and $\phi$ but for the elliptical Petrosian method we use those determined using the second moments within the circularized Petrosian $r_{90}$, the radius containing 90\% of the flux within $2r_P$.  

We have also applied aperture corrections to the photometry, to account for the variation in point spread function between the bandpasses, particularly for GALEX. We do so by using the measured curve-of-growth to predict the aperture correction for an ideal elliptical galaxy, and applying this correction to the real data. For GALEX, these corrections can be of order 30\% to 50\% for galaxies with half-light radii around an arcsecond; for SDSS they are always negligible.

Visual inspection of our targets during the target selection process revealed that the Sersic profile fitting suffered more catastrophic failures than the circular Petrosian profile calculation. A detailed comparison with the \citet{Simard11} two component GIM2D Sersic fits further showed that the NSA's single Sersic \Re~ estimates are systematically overestimated for early-type galaxies (or galaxies with high concentrations) by up to 50\% at high Sersic-n. Adding the elliptical Petrosian fitting maintained the stability of the circularized fits while also measuring axis ratios and position angles, and they do not show systematic differences with the two component Sersic fits. We therefore choose to use elliptical Petrosian $R_e$ and flux measurements throughout. Absolute magnitudes and stellar mass in the NSA, and hence used here, are calculated assuming $H_0 = 100 h km s^{-1} Mpc^{-1}$ with $h$=1.

This extended NSA is designated v1\_0\_1 and is publicly available as part of the SDSS data releases from DR13 onwards.

For the MaNGA selection we limit the extended NSA catalog to those galaxies that lie within the Large Scale Structure mask produced as part of the Data Release Seven NYU Value Added Galaxy Catalog \citep{Blanton05}. This ensures that all targets fall in regions with good SDSS photometry and spectroscopic coverage, and are not close to very bright stars.

\section{Constructing the Targeting Catalog}

Given the parent catalogs defined above, we now discuss the construction of the ``targeting catalog'' that will define the final selection from which the MaNGA targets will be allocated.  We are guided by three requirements: 

\begin{itemize}
\item more than 80\% of the sample should have a specified radius (e.g. 1.5 or 2.5 $R_e$) smaller than our largest IFU bundle\footnote{We define the radius of our hexagonal IFUs to be the radius of the circle that has the same area as the IFU.}.
\item a flat distribution in the stellar mass proxy with a low-mass limit of $\sim$10$^9$ \msun.
\item the selection will only use cuts in redshift that depend on the stellar mass proxy (and one color in the case of the Color-Enhanced supplement). 
\end{itemize}

A summary of the targeting catalog construction is as follows.  We consider three targeting samples.  The goal of the Primary sample is to provide coverage to a radius corresponding to 1.5 $R_e$. The Color-Enhanced supplement (roughly 17\% of the final sample) produces a more uniform coverage in $NUV-i$ color as a function of mass when combined with the Primary sample to form the Primary+ sample. The Secondary sample, designed to yield a sample size that is half of the Primary+ sample, covers larger radii, up to 2.5 $R_e$. For the optimization process that we describe below we only consider the Primary and Secondary samples. The Color-Enhanced supplement, is produced by only slightly widening the Primary sample selection criteria in a color dependent way. That combined with its small size means that it has a negligible effect on the final sample size and S/N distributions and so the optimization based on the Primary and Secondary samples remains valid (see \S\ref{sec:primplus} for a demonstrations of this and a detailed description of the Color-Enhanced selection methodology).   

After choosing the relative proportions of the sub-samples, we first adopt a desired total sky density of potential targets.  This in itself requires an optimization process that balances the efficiency of allocating IFUs, the field-of-view, survey area, and the number and size of IFUs that can be constructed, and trade-offs in S/N, exposure time, spatial resolution and radial coverage.   These are discussed in \S\ref{sec:density}.  Once the desired sky density is defined, we derive stellar mass proxy dependent low- and high-redshift cuts that yield samples that meet the coverage criteria.  These cuts are then optimized to deliver the highest S/N and spatial resolution across the targeting samples (in effect, this means that the lowest redshifts are preferred).  We then ``tile'' the survey---a term that refers to the selection of MaNGA pointings across the sky and the allocation of IFUs to targets.  This allows us to evaluate the final ``observed'' sample that is obtained as well as the frequency of unused or improperly allocated IFU bundles. We repeat the process multiple times under different assumptions for the target density, the minimum and maximum IFU sizes, and the distribution of fabricated IFU sizes to determine the optimal configuration. Further details are given below.

\subsection{Selecting Upper and Lower Redshift Cuts}
\label{sec:zcuts}

 \begin{figure}[h]
 \begin{center}

 \includegraphics[width=7cm]{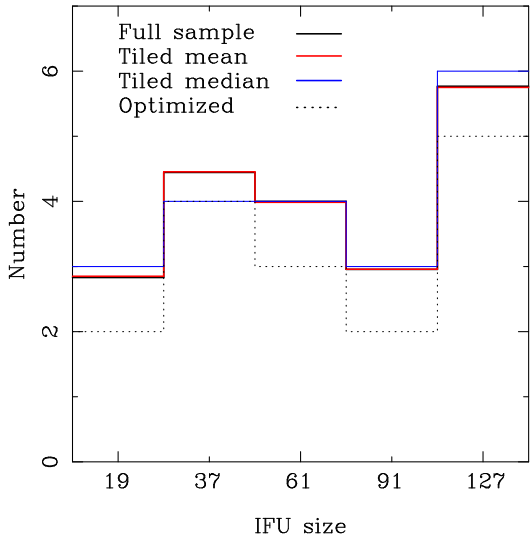}

 \caption{\small The required IFU size distribution to cover the primary sample to 1.5$R_e$ and the Secondary sample to 2.5$R_e$. Galaxies smaller than 19 fibers are assigned to 19-fiber IFUs; galaxies larger than 127 fibers are assigned to 127-fiber IFUs. The solid black histogram indicates the mean size distribution of the whole sample. The red histogram shows the mean size distribution per tile after allocating IFUs using a non-overlapping tiling of the sample. The blue histogram shows the median distribution per tile. All three solid histograms have been normalized to a total of 20 galaxies. The dotted histogram is the optimal IFU size distribution that can fit on the slit. See text for the exact optimization procedure.}
  \label{fig:bunsizedist}
 \end{center}
 \end{figure}

\begin{table*}[ht]
  \begin{center}
    \caption{\label{tab:maxbunsizesamp} The properties of Primary samples designed for differing max IFU sizes.}
	\begin{tabular}{cccccccc}
\\
	\tableline
    	\multicolumn{1}{c}{Max IFU} &
      	\multicolumn{1}{c}{Density} &
      	\multicolumn{1}{c}{N$_{gals}$} &
      	\multicolumn{1}{c}{N$_{IFU}$} &
      	\multicolumn{1}{c}{N$_{plates}$} &
      	\multicolumn{1}{c}{Fraction of} &
      	\multicolumn{1}{c}{Median S/N per} &
      	\multicolumn{1}{c}{Median S/N per} \\
    	\multicolumn{1}{c}{size} &
      	\multicolumn{1}{c}{(deg$^{-2}$log($M_*$)$^{-1}$)} &
      	\multicolumn{1}{c}{} &
      	\multicolumn{1}{c}{} &
      	\multicolumn{1}{c}{} &
      	\multicolumn{1}{c}{unused IFUs} &
      	\multicolumn{1}{c}{kpc$^2$ at 1.5 $R_e$} &
      	\multicolumn{1}{c}{$R_e^2$ at 1.5 $R_e$} \\
	\tableline
91  & 1.2 & 9008 & 22 & 474 & 0.14 & 3.8 & 220\\
127 & 1 & 9006 & 17 &  596 & 0.11 & 4.6 & 268\\
169 & 0.8 & 9005 & 13 &  802 & 0.14 & 5.2 & 304\\
217 & 0.8 & 9003 & 11 &  915 & 0.11 & 5.7 & 344\\

	\tableline
    \end{tabular}
\end{center}
\end{table*}

 Once the desired sky density has been set (see \S\ref{sec:density}), we identify redshift intervals at every stellar mass where $>$80\% of galaxies with that mass have a physical scale (either 1.5 or 2.5 $R_e$) that subtends an angular size that fits within the largest available IFU (for discussion on the maximum IFU size, see \S\ref{sec:maxsize}). There are many such redshift intervals, of course.  By choosing the interval with the lowest redshift we maximize both the spatial resolution (in kpc) and the S/N of the resulting sample, while maintaining both the radial coverage and density criteria.  We impose a hard lower redshift limit of z = 0.01, designed to minimize the distance errors introduced by the local velocity field. This lower redshift limit also has the effect of limiting the main samples to stellar masses larger than $\sim 4 \times 10^8\Msun$.

In practice, we bin the parent catalog into a fine grid in log stellar mass (or absolute magnitude) and redshift. 
For each stellar mass bin we find all the redshift ranges that produce the target density. We then find the lowest redshift range that yields a sample in which 80\% of galaxies can be covered to 1.5 or 2.5 $R_e$ (for the Primary and Secondary samples, respectively). This produces an upper and lower redshift limit for each stellar mass bin.  We then interpolate to find the appropriate redshift thresholds for every stellar mass in the parent catalog.

Figure \ref{fig:smzcuts} shows the results of applying this method under the assumption of a sky density of 1 deg$^{-2}$ log(M$_*$)$^{-1}$ and a maximum IFU size of 127 fibers.  While the upper and lower redshift cuts (top left panel) may look somewhat convoluted, they produce a sample that has a very similar angular size distribution across all stellar masses. 
This means that we probe the same spatial resolution in units of $R_e$ at all masses, although the physical resolution in kpc is mass-dependent. We will return to this point later. We note that the change in the distributions for the highest mass galaxies is unavoidable.  Due to the steepness of the mass function there is a shortage of these galaxies at very low redshifts. 

The bottom panels of Figure \ref{fig:smzcuts} also show that even in a narrow mass and redshift range the galaxies show a wide variation in size. In fact even if we were to look at a single stellar mass (or magnitude) and redshift there is still a significant range in galaxy $R_e$ ({\it rms} $\sim$ 50\%). Therefore, a range in IFU sizes is required to most efficiently observe the sample.

\subsection{Optimizing the IFU Size Distribution}
\label{sec:bunsizedist}

\begin{table*}[ht]
  \begin{center}
    \caption{\label{tab:maxbunsizedist} Optimal IFU size distributions}
	\begin{tabular}{ccccccccc}
\\
	\tableline
    	\multicolumn{1}{c}{Max IFU size} &
      	\multicolumn{1}{c}{N19} &
      	\multicolumn{1}{c}{N37} &
      	\multicolumn{1}{c}{N61} &
      	\multicolumn{1}{c}{N91} &
      	\multicolumn{1}{c}{N127} &
      	\multicolumn{1}{c}{N169} &
      	\multicolumn{1}{c}{N217} &
      	\multicolumn{1}{c}{mean square difference/dof} \\
	\tableline
91  & 4 & 6 & 5 & 7 & 0  & 0 & 0 & 5584\\
127 & 2 & 4 & 4 & 2 & 5 & 0 & 0 & 5794\\
169 & 1 & 2 & 3 & 2 & 1 & 4 & 0 & 7072\\
217 & 1 & 2 & 2 & 1  & 1 & 1 & 3 & 6182\\

	\tableline
    \end{tabular}
\end{center}
\end{table*}

 Physical size and mass are correlated 1:1 (to first order), and we wish to sample a factor of 100 in mass. To match this dynamic range requires either a comparable range in IFU size (for a pure, volume-limited sample), or selecting more massive galaxies at preferentially higher redshift, thus lowering physical resolution in a mass-dependent way. As the dynamic range in IFU size increases the total number of IFUs decreases (for fixed total fiber number). This decrease in the number of IFUs then requires a lower target density, a larger survey footprint and, for a fixed number of targets, shorter exposures.

To properly optimize these trade-offs we need to investigate the final sample properties after a full realization of our targeting and selection algorithms, rather than just analyzing the targeting catalog itself. In particular we must include the effects of the field size (i.e we must tile the survey) and available IFUs. We first tackle the question of how to optimize the IFU size distribution given a maximum and minimum IFU size. In the next section we will discuss how the maximum and minimum size is chosen.

Since our final sample of galaxies will have a range of apparent angular sizes (see Figure \ref{fig:smzcuts}) we would like to have a range of IFU sizes that is able to closely match this distribution. We always wish to observe a galaxy with an IFU that is large enough to reach the desired radius, but, to maximize survey efficiency, we do not wish to use an IFU any larger. If in a given tile\footnote{We adopt the SDSS terminology that defines a pointing of the Sloan 2.5m field-of-view on the sky as a ``tile.''  A given plate is associated with a set of drill holes that locate fibers on specific targets.  Thus, more than one plate can be observed over a given tile.  A full set of tiles, which may also overlap, describes the footprint of the survey.} we have many more targets than available IFUs we can select the galaxies that match our IFU size distribution and thus maximize our survey efficiency. However, if the IFU distribution does not match the underlying galaxy size distribution we will produce a final sample that is biased compared to our input sample. For these reasons we want to carefully select the IFU size distribution that most closely matches the per tile size distribution derived from the targeting catalog. 
    
Figure \ref{fig:bunsizedist} shows the distribution of required IFU sizes for a potential targeting catalog. The construction of this catalog assumed a Primary-to-Secondary ratio of 2 to 1, a maximum IFU size of 127 fibers, and a Primary sample density of 1 deg$^{-2}$ log(M$_*$)$^{-1}$. Galaxies that require an IFU size smaller than 19 fibers are assigned to a 19-fiber IFU and likewise galaxies that require an IFU size larger than 127 fibers are assigned a 127-fiber IFU.  The black histogram in Figure \ref{fig:bunsizedist} shows the size distribution derived from the full targeting catalog, assuming that {\em all} galaxies can be equally well observed.  In fact, galaxies can ``collide'' (if a pair is closely separated only one may be allocated an IFU) and are highly clustered resulting in some regions with more or fewer available targets than MaNGA has IFUs.  

After running a non-overlapping tiling of the targeting catalog (see \S\ref{sec:tiling}), the importance of these effects can be judged by the resulting mean (red histogram) and median (blue histogram) distributions, defined {\em per tile}.  In all cases, the size distributions are scaled to a total number of 20 IFUs, which is the median number of target galaxies per tile in the adopted tiling scheme.  The red or blue histogram in comparison to the black histogram represents two extreme tiling strategies. The non-overlapping tiling is the most efficient in terms of maximizing the number of galaxies observed with a given number of plates while still probing all environments, but makes no attempt at completeness. The distribution in the full targeting catalog represents 100\% completeness while paying no attention to efficiency. Our eventual strategy will be somewhere between the two but we can see that there is practically no difference between the two means (black vs red).

Since we are limited to selecting integer numbers of IFUs, the blue histogram in Figure \ref{fig:bunsizedist}, the median distribution of the non-overlapping tiling, looks to be a good solution. However, these IFUs would require more fibers than can fit on the slit of the BOSS spectrograph even with our minimum acceptable slit spacing \citep[see][]{Drory15}. Therefore, we must find the IFU size distribution that most closely matches these required distributions but requires fewer fibers than can fit on the BOSS spectrograph slit. 

To achieve this optimization we perform an exhaustive search over a large number of IFU size distributions where the number of IFUs of a given size varies from 0 to 8 and where the total slit space consumed is always less than the maximum available. We then calculate the mean square difference between each test IFU size distribution and the actual required IFU size distribution, where both are normalized to a total number of 20 IFUs. We do this both for the full targeting catalog size distribution and for each of the non-overlapping tiles. In the case of the non-overlapping tiles the mean square difference is then summed over all tiles. The {\it best} IFU size distribution is then that which minimizes the mean square difference. 

For the sample described in this section the optimal IFU size distribution is 2,4,3,2,5 for both the tiled and untiled samples. It is shown as the dotted line in Figure \ref{fig:bunsizedist}. We note that a distribution of 2,4,4,2,5 is almost as good a fit and can also be accommodated on the slit. Since it wholly contains the optimal distribution but includes an extra IFU it would be logical to choose this distribution as it will yield a larger final sample that can be reduced to the optimal distribution (2,4,3,2,5) after the observations are completed, if so desired.  

\subsection{Selecting the Maximum and Minimum IFU Size}
\label{sec:maxsize}

 \begin{figure*}[ht]
 \begin{center}

 \includegraphics[width=16cm]{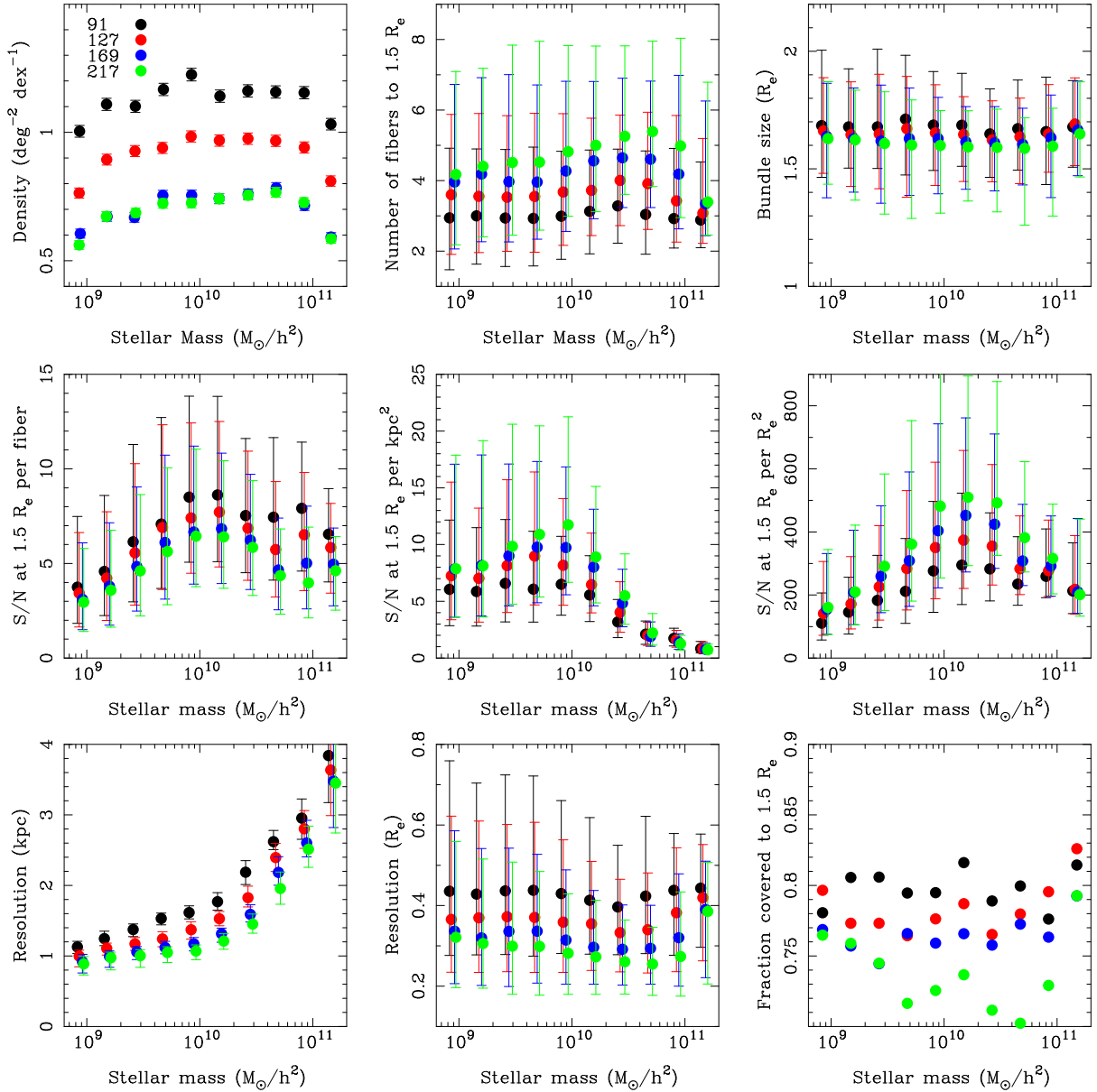}

 \vspace*{0cm}
 \caption{\small Performance comparison of different samples designed with different maximum IFU size. Black, red, blue, and green symbols represent samples with a maximum IFU size of 91, 127, 169, and 217 fibers, respectively. See text for detail.}
  \label{fig:maxbunsize}
 \end{center}
 \end{figure*}

Early work defining the survey and instrument strategy resulted in the definition of a sample that required an IFU size range from 19 to 127 fibers. This size range was determined by a combination of the requirements of the initial sample selection and the properties of the instrument and has been used for the IFU development work. In this section we revisit this choice of IFU size range and investigate if it is optimal.

\subsubsection{Minimum IFU Size}

The choice of the the 19-fiber IFU as the minimum possible size is a simple one. We require at least 3 radial bins for all of our science cases and so smaller IFUs are not worthwhile. Our sample selection methodology (\S\ref{sec:zcuts}) is not directly constrained by the minimum IFU size, but instead maximizes the angular size of the sample. As such, we can make the 19-fiber IFU available to our IFU size distribution optimization procedure (\S\ref{sec:bunsizedist}) and see if it is required for a given sample.

\subsubsection{Maximum IFU Size}

 \begin{figure*}[ht]
 \begin{center}

 \includegraphics[width=16cm,angle=0]{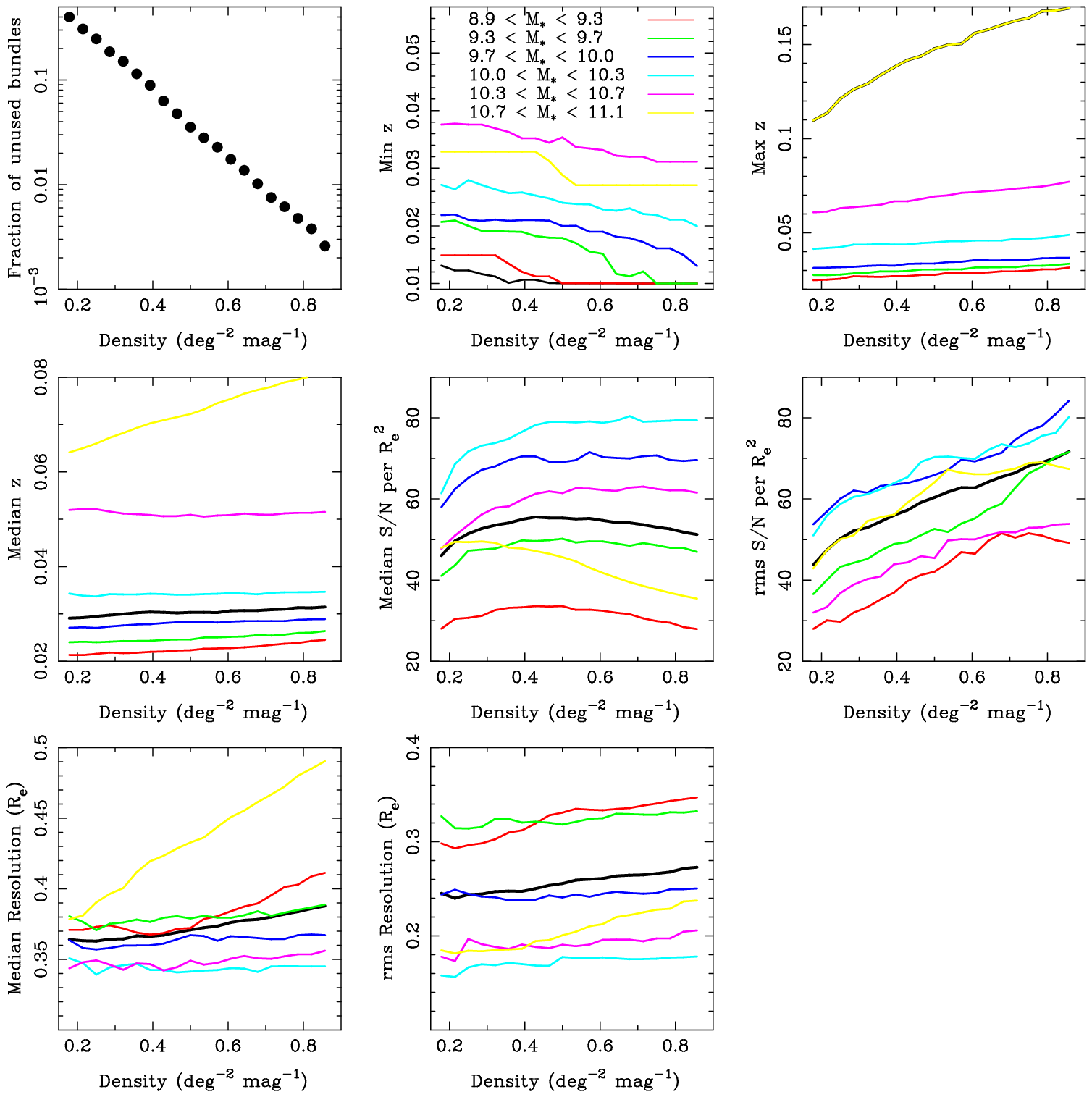}

 \vspace*{0cm}
 \caption{\small A comparison of key properties among samples designed with different target densities per square deg. per magnitude. The top left panel shows the fraction of unused IFUs decrease as we increase the number density of potential targets on the sky. The other panels show various properties as a function of the density. In each panel, colored lines indicate different stellar mass bins and the black line indicates the property for the whole sample. See text for more details.}
  \label{fig:density}
 \end{center}
 \end{figure*}

The choice of a maximum IFU size is somewhat more complex as it enters directly into the determination of the stellar mass dependent redshift cuts that define our samples (\S\ref{sec:zcuts}). A larger maximum IFU size will allow a sample to have a lower average redshift and still be covered to the same physical scale (e.g. 1.5 or 2.5 $R_e$). Clearly selecting a sample at lower redshift will improve the resolution and could potentially increase the S/N. The downside is that an increase in the IFU size reduces the total number of IFUs that will be available, since the slit length and hence number of fibers is fixed. Thus, to achieve the same sample size with fewer IFUs the number of plates observed must increase, and thus with a fixed amount of survey time available the exposure time per plate must be reduced. 

We investigate these tradeoffs by designing a series of samples with different maximum IFU sizes of 91, 127, 169, 217. In each case we design Primary and Secondary samples where 80\% of the galaxies are covered to 1.5 and 2.5 $R_e$ respectively by the maximum IFU size. We choose the target densities of each sample so that when they are tiled each sample has the same fraction of IFUs that are unused due to tiles with too few targets on them. For each sample, we optimize the IFU size distribution using the procedure described in \S\ref{sec:bunsizedist} utilizing a full non-overlapping tiling, with the results shown in Table \ref{tab:maxbunsizedist}.

We then tile each of these samples using the optimized IFU distribution and a non-overlapping tiling, selecting the number of tiles required to produce a sample of 9,000 galaxies. For each tile, galaxies are selected to match the available IFU sizes. If there are more IFUs of a given size than galaxies of that size, they are allocated to galaxies that cannot be allocated an IFU of the correct size (see \ref{sec:allocation} for more details of the allocation process).

Table \ref{tab:maxbunsizesamp} gives some of the properties of these samples. Details of the performance of the Primary sample are shown in 
Figure \ref{fig:maxbunsize}. The S/N values in the table and plots are calculated using an exposure time inversely proportional to the required number of plates. We assume an exposure time of 3 hours for the sample designed for a maximum IFU size of 127 fibers and scale the exposure time for the other samples accordingly to ensure the same total survey time. 

The top row of panels in Figure \ref{fig:maxbunsize} simply show that the Primary samples are performing as designed. The center row compares the S/N properties and the bottom row the resolution and coverage.  We assume an effective angular resolution of 2.5\arcsec\ based on expectations for the reconstructed data cubes.  The center left panel shows that the median S/N per fiber at 1.5 $R_e$ decreases as the maximum IFU size increases. This S/N decrease is simply due to the decrease in exposure time per plate, required since we must observe more plates to achieve the same final sample size with fewer IFUs per plate.
A better representation of the S/N is given by the two other panels in the center row, which show the S/N at 1.5 $R_e$ per kpc$^2$ and per $R_e^2$ respectively\footnote{The S/N per $R_e^2$ at 1.5 $R_e$ is a useful if perhaps unusual metric.  To compute it, we consider the integrated S/N in a small annulus (or fiber) positioned at 1.5 $R_e$.  We then divide by the area of that annulus in units of $R_e^2$. This naturally accounts for the fact that the angle subtended by $R_e$ on the sky (e.g., in arcseconds) depends on the galaxy's intrinsic size and its redshift. Furthermore, this S/N metric is appropriate for addressing the fidelity of measurements of both kinematics and compositional properties that scale with $R_e$.}. The opposite trend is now apparent, with the S/N increasing as the maximum IFU size increases, since each kpc or unit of $R_e$ covers more fibers at lower redshift. This trend is confirmed by the median S/N values for the whole of each sample given in Table \ref{tab:maxbunsizesamp}. The largest fractional S/N increase occurs when increasing the maximum IFU size from 91 to 127 fibers, with the relative size of the increase diminishing as the maximum IFU size increases beyond 127.

As we allow larger IFUs to be considered the resolution increases as expected. Once again the biggest improvement is seen when going from a maximum IFU size of 91 to 127 fibers. This trend is not surprising since we increase the radius by one fiber each time and so there is a larger fractional increase at smaller IFU sizes. What is less obvious is the strong stellar mass dependence to the gain in resolution, with the largest effect occurring at stellar masses of a few 10$^{10}$ \msun. This reflects the fact that galaxies in this stellar mass range have the largest mean angular sizes because they are intrinsically larger than low mass galaxies, but the turnover of the mass function means higher-mass galaxies must be targeted at increasingly more distant redshifts.

The final panel (bottom right) shows the fraction of galaxies that are covered to at least 1.5 $R_e$ after tiling. There is a general slow reduction in this fraction as the maximum IFU size increases, but a sudden fall from 169 to 217. The fraction of the IFU complement made up by the largest IFU bundle size does decrease as the maximum allowed size increases, leading to a reduction in the fraction covered to the target radius. There are also fewer galaxies per plate for a given IFU size making it harder to match the galaxies to the IFUs. This could be mitigated with a higher density sample but there would be a subsequent loss of resolution (see \S\ref{sec:density} below). 

While increasing the maximum IFU size from 127 to 169 will lead to some gain in S/N and resolution (it should be noted that the same effect causes both to increase) with a moderate loss of coverage fraction, there are some more practical disadvantages to larger IFU sizes. Since larger IFUs require more slit space, the total number per plate decreases. This results in a larger number of plates required for the same sample size and hence a higher plate production cost (30\% more plates $\sim$\$100k). Furthermore, building and testing an IFU larger than the 127 fiber IFU would have required further development again increasing costs and placing the schedule at risk. We therefore settled on a final IFU size range of 19 to 127 fibers.

\subsection{Choosing the Sample Density}
\label{sec:density}

It is possible to construct targeting catalogs meeting our science specifications with different number densities on the sky. Given that the spatial density of galaxies varies on the sky, a higher density of potential targets allows more efficient tiling and more efficient use of IFUs.
A higher density can be achieved by widening the redshift intervals at a given
stellar mass. 
While the average redshift would remain roughly constant, as required by the
desire for a constant angular coverage, a wider redshift interval would result in 
a wider spread in angular sizes, increasing the tension between the dynamic range of 
galaxy sizes and the dynamic range of IFU sizes. 
In addition, as the desired sky density is increased, if a hard lower redshift limit (e.g. $z=0.01$) is reached or there are too few massive galaxies at low-$z$,
the average redshift must be increased, resulting in poorer spatial resolution and total S/N.

Conversely, higher density samples have the advantage of requiring fewer plates with unused IFUs allowing us to reach the desired sample size of the main samples with fewer plates. Since our total time is fixed we may increase the exposure time and thus potentially the S/N.

Overall, input samples with higher density may have a slightly
higher S/N (or more galaxies) but at a potential cost of lower spatial
resolution and greater sample variance in both spatial resolution and
S/N.

To investigate this trade off, we generate several samples using the same procedure 
as described in \S\ref{sec:zcuts} with a large range in sky density and \Mi~as our stellar mass proxy. These
samples are then tiled using a 2,4,4,2,5 IFU size distribution
and a non-overlapping tiling, adding tiles until a sample of 10,000
galaxies is reached. A Secondary sample is also included which has
50\% of the density of the Primary sample.

Figure \ref{fig:density} shows the properties of these Primary input samples constructed from targeting catalogs with varying densities. The top left panel simply shows how the fraction of unused IFUs depends on density, showing a rapid decline as density increases which asymptotes to zero at high densities as expected. Reading from left to right and top to bottom, the next three panels show the density dependence of minimum, maximum and median redshift for all galaxies (black) and split by stellar mass (color). One can see that as the density increases the minimum and maximum redshifts diverge as expected. It is also evident that the median redshift increases little, except where z$_{min}$(\Mi) hits a limit which is most evident for the highest and lowest stellar mass bins.

The final four panels show the median and {\it rms} of the S/N and
resolution respectively. The S/N is determined in a fiber at 1.5 $R_e$ and is divided by the area of the fiber in units of $R_e^2$. Likewise the resolution is given in units of $R_e$. For each density the S/N is scaled by the square root of the relative number of plates required to reach 10,000 galaxies, representing the change in plate exposure time available in a fixed duration survey. One can see that as the density is increased the median S/N begins to increase before it reaches a
plateau or turns over and begins to decrease. This turnover happens
most rapidly for the lowest and highest mass samples reflecting
their larger changes in median redshift, which counteracts the
increased exposure time and decreases the S/N. The median resolution
again shows the largest trend for the highest and lowest mass samples
as it simply tracks the median redshift and thus increases (degrades) with density. In both cases the {\it rms} increases with density,
reflecting the widening high and low redshift limits. 

Figure \ref{fig:density} makes it clear that we do not wish to target samples with densities $>$ 0.6 deg$^{-2}$ mag$^{-1}$ since the S/N has either flattened off or is declining at this point while the resolution gets poorer and the scatter in both quantities increases. However, at densities below this there is a trade-off between S/N and resolution. A density of 0.53 deg$^{-2}$ mag$^{-1}$ maximizes the overall median S/N while only reducing the median resolution by 1\% over the whole sample and produces similar results for the individual stellar mass bins with the exception of the highest stellar masses. We therefore select this density for the Primary sample.

\section{Final Targeting Catalogs}
\label{sec:finalsamp}

We have described above our procedure and optimization strategy for constructing targeting catalogs for our Primary and Secondary samples. We have decided to allow IFU sizes of 19, 37, 61, 91, and 127 fibers and a density for the Primary+ sample of 0.53 deg$^{-2}$ mag$^{-1}$. If we wish to have a Secondary sample of 50\% the size of the Primary+ sample we would require a density of 0.37 deg$^{-2}$ mag$^{-1}$. This is not simply a factor of two lower than the Primary+ density since the \Mi\ completeness limit of the extended NSA (Equation \ref{eq:zcomp} below) means that we cover a narrower \Mi\ range in the Secondary sample (\Mih\ $\lesssim -18$) than in the Primary sample (\Mih\ $\lesssim -19$). However, we have chosen to design the Secondary sample with a somewhat higher density of 0.5 deg$^{-2}$ mag$^{-1}$. The higher density increases the redshift range somewhat making the sample less sensitive to cosmic variance and thus reduces the plate to plate variation in the number of targets making it easier to allocate IFUs to galaxies with the correct size. To maintain the desired 2 to 1 ratio of Primary+ to Secondary galaxies we down-sample the Secondary sample to a density of 0.37 deg$^{-2}$ mag$^{-1}$ during the IFU allocation process.

\subsection{Simplifying the Selection Function}
\label{sec:selfn}

\begin{figure*}[ht]
 \begin{center}

 \includegraphics[width=16cm]{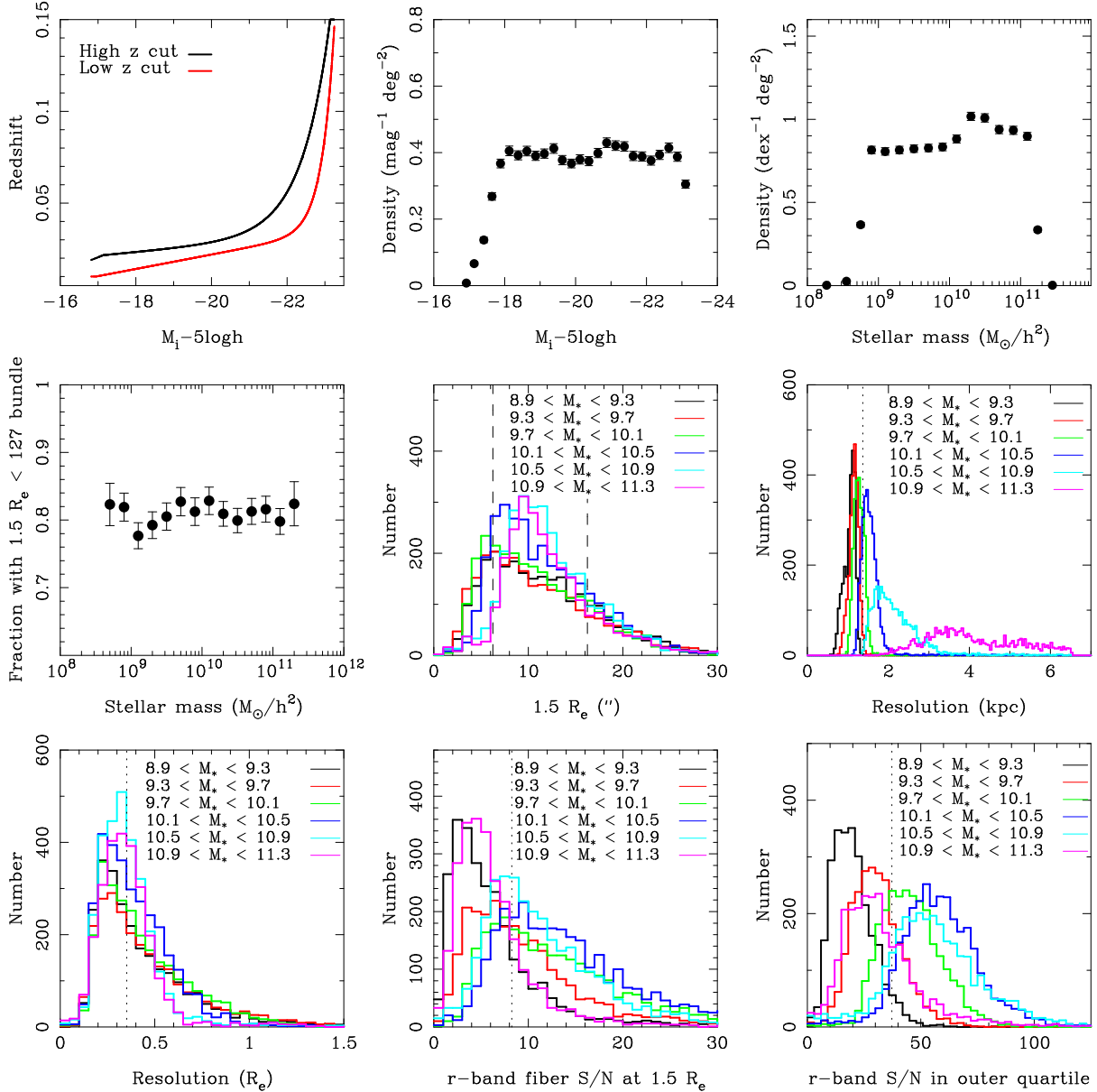}

 \vspace*{0cm}
 \caption{\small The selection (first three panels) and detailed properties (other panels) of the final Primary sample. Colored histograms indicate different stellar mass bins. See text for detail.}
  \label{fig:finalpriprop}
 \end{center}
 \end{figure*}

Since we have elected to design samples that have flat number densities as a function of a stellar mass proxy (and are flattened in color as well in the case of the Color-Enhanced supplement) we will need to correct for this imposed selection function for any statistical analysis of the MaNGA sample. Since the only selection we impose is an upper and lower redshift limit as a function of our stellar mass proxy (or color and mass for the Color-Enhanced supplement) we can exactly define the volume over which any galaxy in our samples could have been selected. This allows the easy calculation of a Vmax weight for every galaxy in the sample enabling the sample to be corrected back to the volume limited case (see \S\ref{sec:vmax} for details). 

However, this Vmax is only perfectly defined in the case where there are no errors on the selection parameter (e.g. mass, magnitude, color etc). Larger errors in the selection parameter, or the combination of errors on multiple selection parameters, translates to a larger error in the weight calculation, which would then translate into errors in any derived relations defined from the MaNGA sample. This could be particularly troubling if the error in the weight ended up correlating with an interesting derived parameter from the MaNGA IFU data.     

For this reason we have elected to define the Primary and Secondary samples using just \Mi~rather than a full estimation of the stellar mass, and the Color-Enhanced supplement using $NUV-i$ color rather than an estimate of the star formation rate. This is also the reason why we have separated the Primary and Color-Enhanced samples as we have, rather than making the Primary sample fully flattened in both a mass and a SFR proxy. Such a split allows the use of just the Primary sample for analyses that may be particularly sensitive to errors in the weights, but still enables analyses that will need better statistics in the lower density regions of the SFR-Mass plane.

Flattening the density in \Mi\ rather than mass has the disadvantage that we end up with a significant number of low mass ($10^8- 10^9\Msun$) blue galaxies in the sample. We therefore choose to remove these by including a color dependent absolute magnitude limit at the faint end. While this does not produce a hard cut in stellar mass, it does remove most of the very low mass blue galaxies. These cuts are defined as

\begin{equation}
	 g-r > 0.4 (M_r - 5logh) + 7.4  
\end{equation}
and
\begin{equation}
	 g-r > 0.28 (M_r - 5logh) + 5.6 
\end{equation}
for the Primary and Secondary samples respectively, where $g-r$ is the k-corrected color at redshift zero.

The final simplification we make is to define the upper and lower redshift limits for the Primary and Secondary samples using a functional form rather than an interpolation between narrow redshift bins. This is largely done for ease of communication and reproduction and only results in a minor change in the sample properties and minimal impact on sample performance. These limits are defined by the following functional form

\begin{equation}
\label{eq:zlims}
  z_{lim} = (A + B(M_i - 5logh)) (1+exp[C(M_i-5logh -D)])
\end{equation}

with parameters $A, B, C$ and $D$ for the lower and upper redshift limits of each sample given in Table \ref{tab:zlims}.   

\begin{table}[ht]

    \caption{{\label{tab:zlims}\small The fit parameters for the functional form (Equation \ref{eq:zlims}) of the \Mih\ dependent upper and lower redshift limits used to define the Primary and Secondary samples.}}
  \begin{center}
	\begin{tabular}{lcccc}
	\tableline
    	\multicolumn{1}{l}{Redshift Limit} &
      	\multicolumn{1}{c}{A} &
      	\multicolumn{1}{c}{B} &
      	\multicolumn{1}{c}{C} &
      	\multicolumn{1}{c}{D} \\
	\tableline

Primary lower & -0.056597  & -0.0039264   &  -2.9119   & -22.8476 \\
Primary upper & -0.011377  & -0.0019220  &  -1.2924  & -22.1610 \\
Secondary lower & -0.056463  & -0.0048895   &  -1.3773   & -22.3851 \\
Secondary upper & -0.048010  & -0.0046639   &  -1.3719   & -22.3225 \\

	\tableline
    \end{tabular}
\end{center}
\end{table}

We also include a completeness cut required as a result of the magnitude limit of the input catalog where 

\begin{eqnarray}
\label{eq:zcomp}
 \nonumber z &<& -0.9335 - 0.1839 (M_i-5logh) - 0.01222 (M_i-5logh)^2\\
      &-& 2.7668\times10^{-4} (M_i-5logh)^3.
\end{eqnarray}

\subsection{Final Sample Properties}

\begin{figure*}[ht]
 \begin{center}

 \includegraphics[width=16cm]{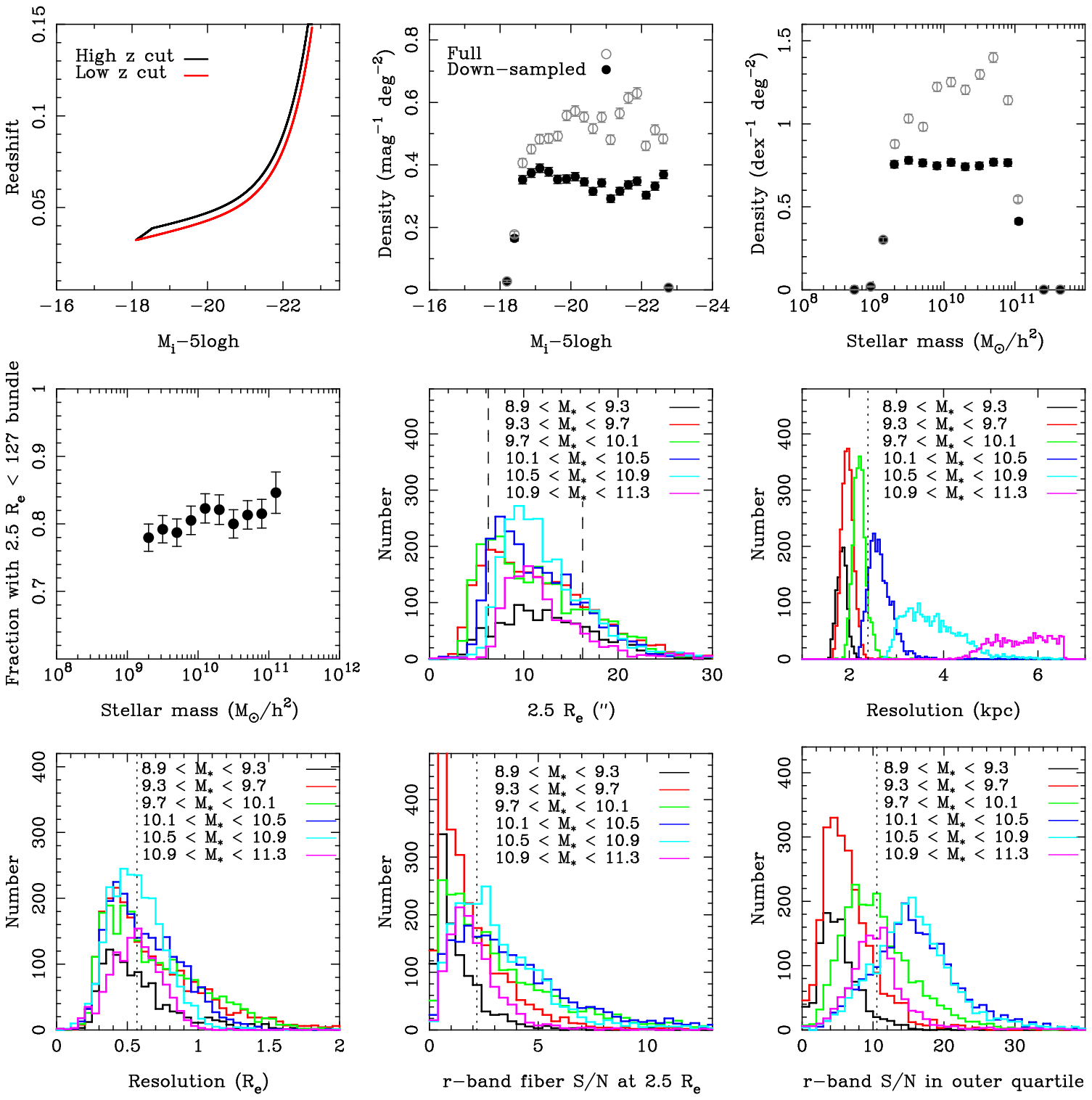}

 \vspace*{0cm}
 \caption{\small The selection (first three panels) and detailed properties (other panels) of the final Secondary sample. Colored histograms indicate different stellar mass bins. See text for detail.}
  \label{fig:finalsecprop}
 \end{center}
 \end{figure*}

Applying our sample design procedure (\S\ref{sec:zcuts}) using \Mi~as the mass proxy, and defining Primary and Secondary samples that provide coverage to 1.5 and 2.5 $R_e$ for 80\% of potential targets results in the \Mi~dependent redshift cuts shown in Figures \ref{fig:finalpriprop} and \ref{fig:finalsecprop}. Applying the IFU size distribution optimization technique (\S\ref{sec:bunsizedist}) yields a preferred distribution of 2, 4, 4, 2, 5 for a total of 17 IFUs per plate. It is worth noting here that varying the target density has little effect on the optimal IFU size distribution as long as the target radius (i.e., 1.5 or 2.5 $R_e$) and maximum IFU size remain the same. This means that including the color enhanced sample or adjusting our sample selection in the future, for example by changing the relative numbers of Primary and Secondary targets, will not result in a loss of efficiency or introduce a bias from the IFU allocation.
  
\subsection{Primary Sample Properties}

\begin{figure*}[ht]
 \begin{center}

 \includegraphics[width=16cm]{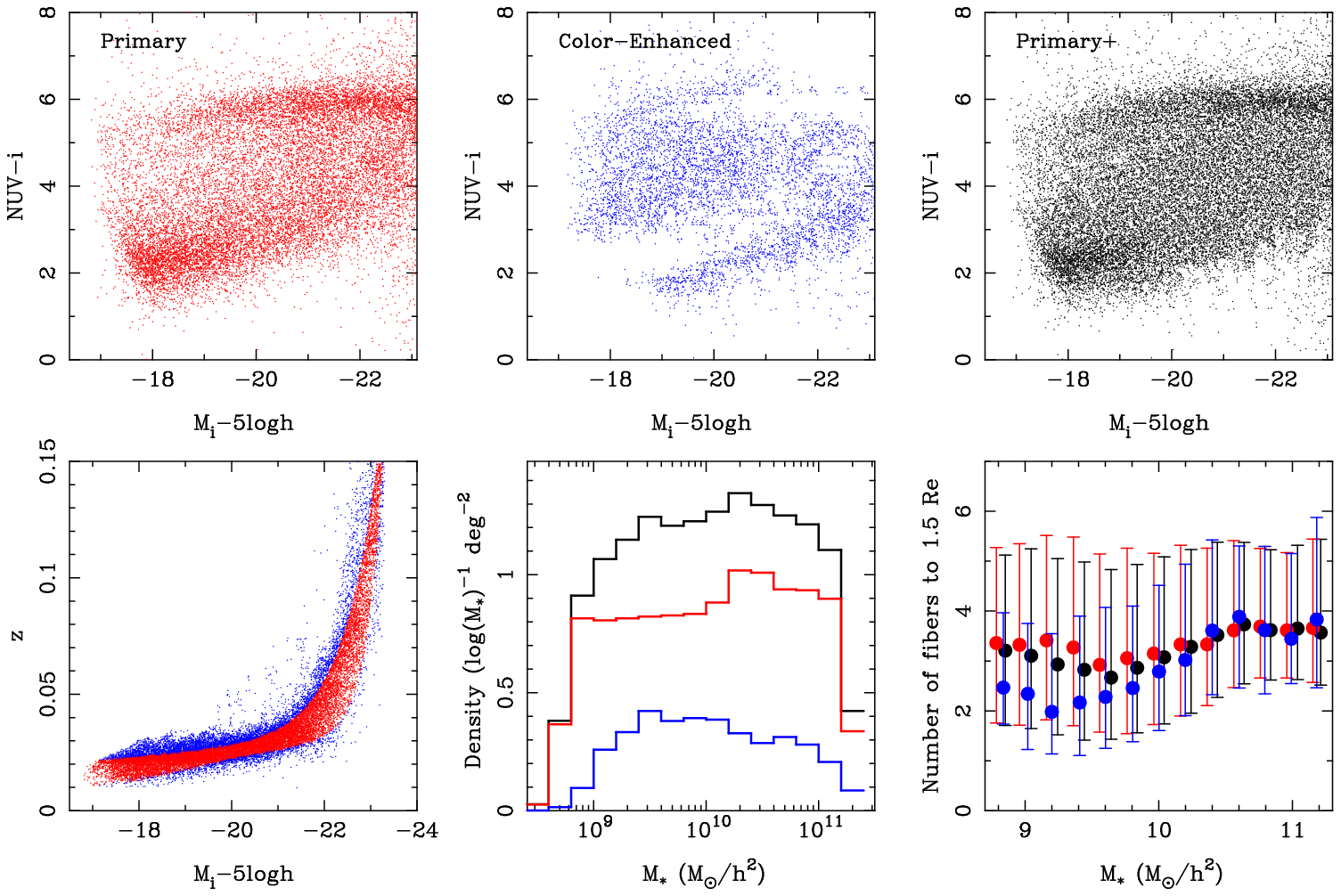}

 \vspace*{0cm}
 \caption{\small The design of the Color-Enhanced supplement. In all panels the Primary sample is shown in red, the Color-Enhanced supplement in blue and the combination (Primary+ sample) in black.  The top row shows from left to right the $NUV-i$ vs $i$ distribution for the Primary, the Color-Enhanced and combination of the two respectively. The Color-Enhanced supplement fills in the regions of this plane that are sparsely populated in the Primary sample. The bottom left panel shows the distribution in the \Mi-redshift plane for the Primary and Primary+ samples. The Primary+ sample includes galaxies at both higher and low redshifts than the Primary sample. The bottom center panel shows the density distribution of the two samples and the combination of the two, as a function of stellar mass. The roughly flat dependence of density on mass is maintained. The bottom right panel shows the median angular size of the galaxies (in units of fiber diameter) for the two samples and their combination. The error bars show the 20th and 80th percentiles. At high masses we are typically adding large galaxies with the opposite being true at small masses. The overall size distribution remains roughly unchanged when adding the Color-Enhanced supplement to make the Primary+ sample. }
  \label{fig:primplus}
 \end{center}
 \end{figure*}

Figure \ref{fig:finalpriprop} shows the properties of the Primary input sample from which actual MaNGA targets will be selected. The top row of panels show the redshift cuts and the density distribution as a function of both \Mi~and \Mstar. The center left panel shows the fraction of targets that can be covered to 1.5\Re\ with a 127 fiber IFU and the center panel the angular size distributions in six equal bins in stellar mass\footnote{We use \Mstar\ rather than \Mi\ as it is the underlying physical parameter of interest.} with vertical dashed lines showing the size of the 19 and 127 fiber IFUs. We can see from these panels that the redshift cuts are doing an excellent job of achieving the desired flat \Mi\ distribution, an approximately flat \Mstar\ distribution and an even angular coverage. All galaxies, irrespective of their mass, have very similar angular size distributions, meaning that they will be covered by the same range in IFU sizes. 80.1\% of the sample have 1.5 \Re\ smaller than the radius of the 127 fiber IFU with just 3\% requiring an IFU smaller than the 19 fiber IFU to reach 1.5 \Re. The introduction of the functional forms for the high and low redshift selection cuts results in much smoother cuts at the expense of some minor variation in the density (top middle panel) and coverage fraction (center left panel).

The remaining panels of Figure \ref{fig:finalpriprop} show distributions of scale and S/N. These are again shown in six mass bins but in these panels the vertical dotted line shows the median for the full sample. The physical resolution in kpc (center row, right panel) depends strongly on stellar mass, especially at high masses, as a result of the typical redshift increasing as the mass increases. For masses $< 10^{10} \Msun$ the median resolution is 1.2 kpc, which increases rapidly to a median of 3.85 kpc for the highest mass bin. The median for the full sample is 1.37 kpc. For the most massive galaxies ($>10^{10.9} \Msun$) there is very little hope for improving the resolution since there are so few of them at low redshifts. For galaxies with 10$^{10} < M_* < 10^{11}$ there are lower redshift galaxies which would yield higher resolutions but these are just too large to be covered to 1.5 \Re. We note that some of these galaxies will be included in an ancillary sample.

When one switches to assessing resolution in terms of \Re\ (bottom left panel) then the sample produces distributions that are almost independent of stellar mass. The median resolution for the full sample is 0.35 \Re. 

The bottom center and right panels give an indication of the $r$-band S/N we can expect to achieve in a typical 3 hour exposure and how it depends on stellar mass. The bottom center panel shows the r-band S/N per fiber at 1.5 \Re\ and the bottom right panel the total S/N in an elliptical annulus (following the ellipticity of the galaxy) covering the outer quartile of the IFU. Typically this annulus will cover 1.35-1.8 \Re. The S/N is lowest for the lowest mass galaxies and increases with mass until the highest masses, where it again decreases. These trends are simply the result of the intrinsic surface brightness distribution of the galaxy population. The medians of the S/N for the full sample at 1.5 $R_e$ are 8.3 per fiber and 37.3 in the outer quartile annulus. Note here that the S/N refers to the spectral S/N per pixel ($10^{-4}$ in log wavelength in \AA) in the r-band.
\\

\subsection{Secondary Sample Properties}
\label{sec:secprop}

Figure \ref{fig:finalsecprop} shows the properties of the Secondary sample in the same manner as for the Primary sample above. The \Mi\ completeness cut Equation \ref{eq:zcomp} is clearly visible in the top left panel as a diagonal cutoff in the high redshift selection cut at faint magnitudes. This reduces the overall range in \Mi\ and hence \Mstar\ sampled compared to the Primary sample, limiting the Secondary sample to \Mstar\ $> 2 \times 10^9$. The other two plots on the top row of the figure again show the density of targets as a function of \Mi\ and \Mstar\ but unlike the Primary sample there are now two density distributions on each plot. The open symbols show the density for the full Secondary sample and the filled symbols after it has been down-sampled to a target density half of that of the Primary sample. Our intention was that this down-sampling would be purely random but in error we continued to use an earlier routine which down-sampled to produce a density exactly flat with stellar mass, as can be seen in the figure. The effect of this error is to add an additional weak dependence on stellar mass to the Secondary selection, which needs to be accounted for in any statistical analysis of the MaNGA sample along side the other selection criteria (see \S\ref{sec:vmax}). Since the required correction is well determined and barely more complex than was already required, and that we had already observed for approximately two years when this error was discovered, we decided that overall it was simpler to continue with this mass dependent down-sampling of the Secondary sample for the full duration  of the survey.

As for the Primary sample these redshift cuts once again meet our IFU coverage target with 80.7\% of the galaxies having 2.5 \Re\ less than the radius of the 127 fiber IFU, and with just 1\% having 2.5 \Re\ smaller than the 19 fiber IFU. Again the angular size distributions are largely independent of stellar mass.

To achieve such coverage one must select a sample at higher redshift than the Primary sample. This has obvious consequences for both the resolution and S/N. The median resolution is a factor of 1.7 poorer for the Secondary sample compared to the Primary sample, although this is partly due to the Secondary sample being restricted to higher masses as a result of the completeness limits. The S/N at 2.5 \Re\ (the edge of the IFU) is low with a median S/N of just 2.3 per fiber and 11.4 in the outer quartile annulus. The aim of this sample is not to study continuum properties at 2.5 \Re\ but those of emission lines, which will naturally have much higher S/N. Even so the expected S/N in the outer annuli should be sufficient for some absorption line studies, and can be increased further by stacking multiple galaxies.

\subsection{Color-Enhanced and Primary+ Samples}
\label{sec:primplus}

\begin{figure}[h]
 \begin{center}

 \includegraphics[width=8cm]{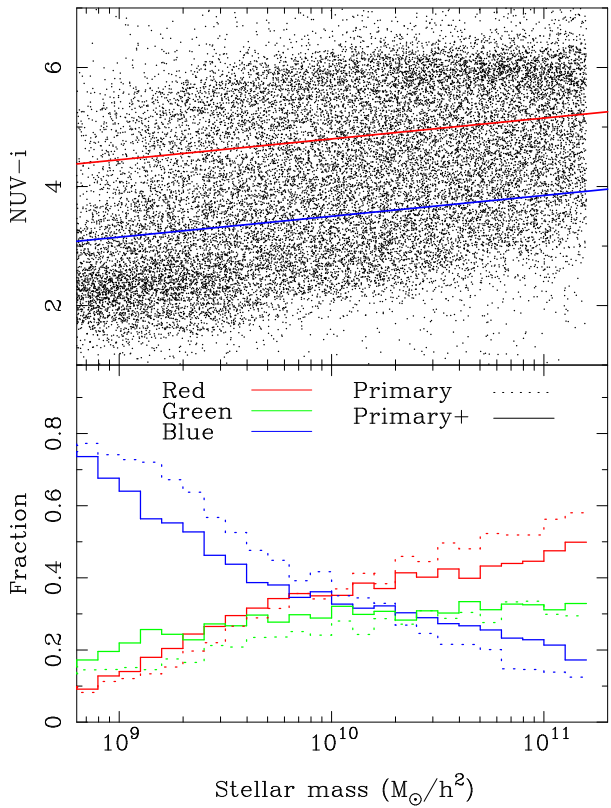}

 \vspace*{0cm}
 \caption{\small Top: The distribution of the Primary+ sample in the $NUV-i$ stellar mass plane. The red and blue lines are used in the bottom panel to divide galaxies into red, green and blue. Bottom: The fraction of red, green and blue galaxies as a function of stellar mass for the Primary (dotted) and Primary+ (solid) samples. The inclusion of the Color-Enhanced supplement in the Primary+ sample flattens these distributions and increases the numbers of rarer galaxies.}
  \label{fig:fraccolsm}
 \end{center}
 \end{figure}

While there are many attractive reasons to simply select a sample that is flat in \Mi, the resulting sample is dominated by red galaxies at high masses and blue galaxies at low masses (see Figure~\ref{fig:fraccolsm}).  A number of MaNGA's primary science drivers concentrate on either red or blue (early or late type) galaxies, and we wish to study how their properties depend on stellar (or dynamical) mass. To improve the statistics for such samples we have designed a Color-Enhanced supplement, which increases the number of galaxies in underpopulated regions of the color ($NUV-i$) versus
magnitude (\Mi) plane. We chose to use $NUV-i$ rather than another color combination because of the wide dynamic range in color it provides. Even though the $NUV$ fluxes have larger absolute errors than say the SDSS $g$-band the much larger dynamic range means that galaxies can still be better separated than if we'd used $g-i$ for example. The Color-Enhanced supplement is designed to be observed to 1.5 \Re\ and will be combined with the Primary sample to form the Primary+ sample.  

We construct the Color-Enhanced supplement by considering regions of the color--magnitude plane for which the number density of galaxies is $<36\%$ of the peak density, which occurs at the luminous end of the red sequence.  We calculate these number densities over regions 0.1~mag wide in \Mi\ and 0.2~mag wide in $NUV-i$ across a range in absolute magnitude ($-16.5>M_i-5logh>-23.5$) and color ($1.6<NUV-i<6.4$).  When considering the bluest and reddest color bins, we also include galaxies with $NUV-i<1.6$ and $NUV-i>6.2$, respectively.  For each bin in the color-magnitude plane we expand the redshift limits as defined by the Primary sample to include more galaxies using the following procedure: \\

\begin{enumerate}

\item We expand the redshift limits until the number density for the Primary+ sample in that bin is 36\% of the peak density, or the number density is increased by a factor of three, or we hit the limiting redshift range of the parent catalog (0.01 $<z<$ 0.15).

\item If fewer than 50\% of the newly added color-enhanced galaxies in the bin have 1.5$R_e$ lying within our IFU size range we reduce the redshift limits of that bin until $\geq 50$ do have 1.5$R_e$ lying within our IFU size range, or until half of the candidate Color-Enhanced galaxies have been removed.

\end{enumerate}

These criteria strike a balance between increasing the number density and ensuring coverage to the target radius.

As for the Primary and Secondary samples we always apply a minimum redshift limit of z = 0.01 and a maximum redshift limit below the \Mi\ completeness limit (Equation \ref{eq:zcomp}). 

The demographics of the full Primary+ sample are illustrated in Figure \ref{fig:primplus}. The top row shows from left to right the distribution in the NUV-$i$ vs \Mi\ plane of the Primary, Color-Enhanced and Primary+ (the Primary plus the Color-Enhanced) samples. The Color-Enhanced selection is mainly adding galaxies in the {\it green valley} and in the faint end of the red sequence and the bright end of the blue cloud. The distribution of the Color-Enhanced supplement in the redshift-\Mi\ plane, along with that of the Primary sample is shown in the bottom left panel and the number density of the Primary, Color-Enhanced and Primary+ samples in the bottom middle panel. The bottom right panel shows that the median angular size of the galaxies in units of fiber diameter, with the error bars showing the 20th and 80th percentiles. The Color-Enhanced supplement typically adds galaxies with a smaller angular size (as a result of their higher redshift) than in the Primary sample at low masses and galaxies of a similar size, but with a slightly larger spread at higher masses. This actually results in a slight increase in the fraction of galaxies in the Primary+ sample that can be covered by the largest IFUs to 1.5 \Re\ but a slight decrease in the average spatial resolution. 

Figure \ref{fig:fraccolsm} shows the performance of the Primary+ sample in flattening the color-mass distribution of galaxies. The top panel shows the color-mass distribution along with two lines designed to split the sample into red-sequence, green valley and blue cloud galaxies. The bottom panel shows how the fraction of each of these three galaxy classes depends of stellar mass for both the Primary and Primary+ samples. While there are still more low mass blue galaxies and high mass red galaxies in the Primary+ sample, the trends of red and blue fraction with stellar mass have been flattened by the addition of the Color-Enhanced supplement and the fraction of {\it green valley} galaxies increased. 

\section{Tiling the Survey}
\label{sec:tiling}

\begin{figure*}[ht]
 \begin{center}

 \includegraphics[width=8.8cm]{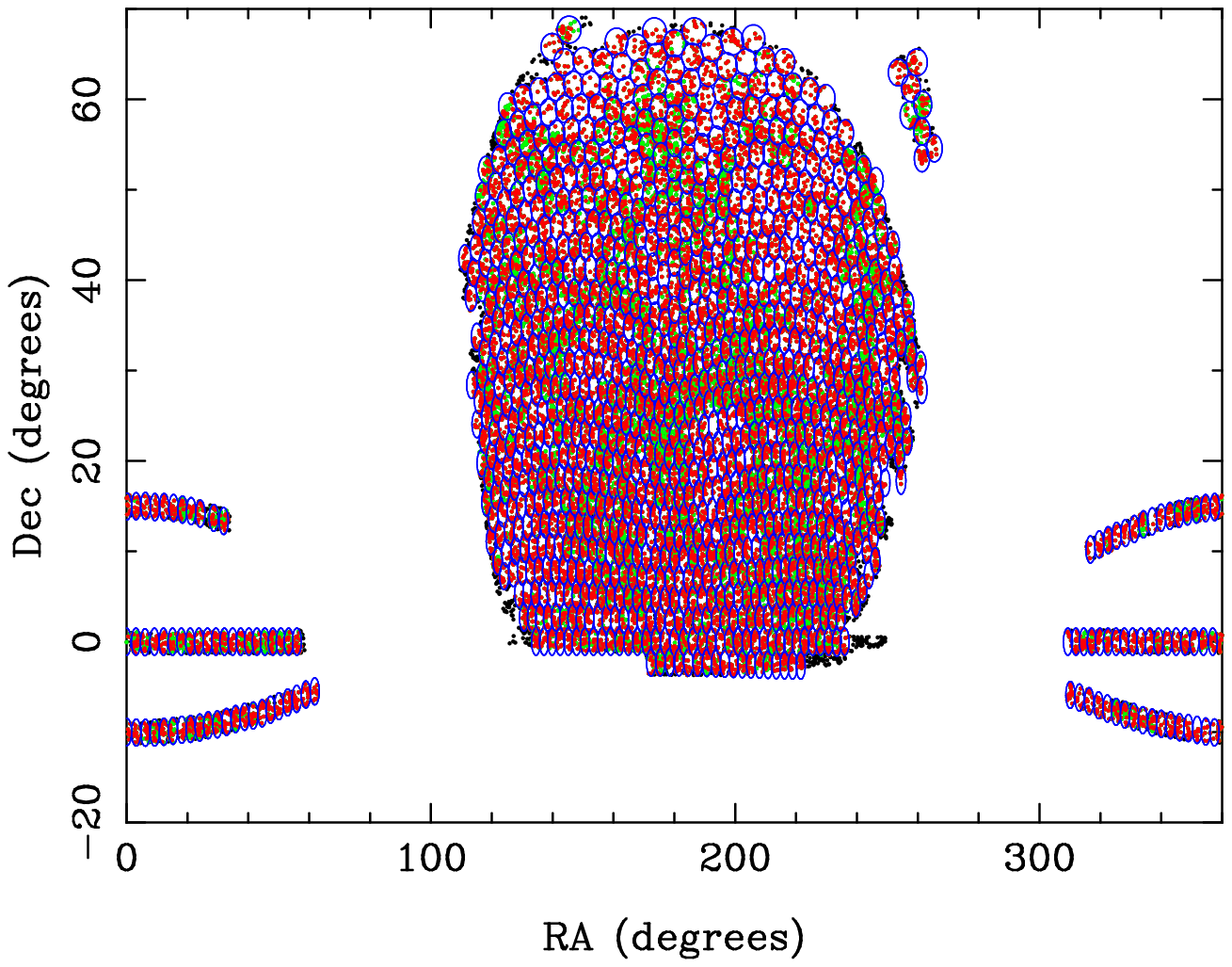}
 \includegraphics[width=8.8cm]{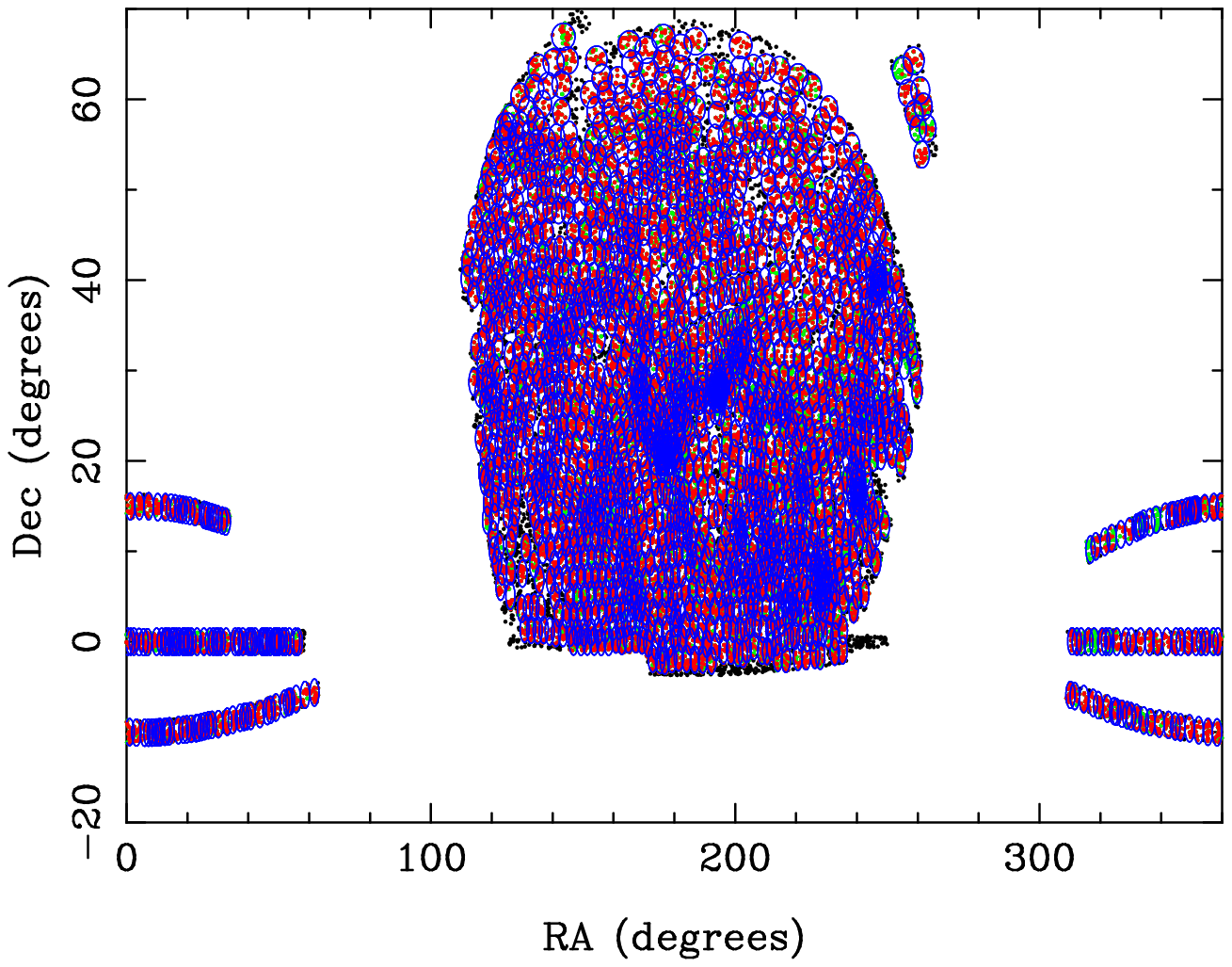}

 \vspace*{0cm}
 \caption{\small An illustration of two potential tiling strategies. The plot on the left shows a non-overlapping tiling and the plot on the right shows an adaptive overlapping tiling designed to produce a more even completeness irrespective of target density. The black points show the positions of galaxies in a our final Primary+ and Secondary targeting catalogs. The blue ellipses show the outlines of the tiles. The green points show those galaxies that lie within the footprint of these tiles and the red points those galaxies that were assigned IFUs}
  \label{fig:tiling}
 \end{center}
 \end{figure*}

In order to execute the survey we must decide how we will cover the area spanned by the targeting catalog and allocate IFUs to potential targets.  We refer to this process as tiling.

\subsection{Pointing strategy}

We first discuss how to divide the full area of the input samples into 7 deg$^2$ circular tiles, each representing the footprint of a single plate. The degree to which these tiles are allowed to overlap, or indeed repeat, represents a trade-off between efficiency and completeness. 

The left panel of Figure \ref{fig:tiling} shows the distribution of galaxies in our Primary+ and Secondary samples over the full SDSS DR7 footprint, along with an example tiling designed to have non-overlapping tiles. Looking at this figure it is apparent that the number of targets per tile varies considerably; the mean number of potential Primary+ and Secondary MaNGA targets is 27.0 per tile, with an $rms$ of 15.9, a minimum of 2, and a maximum of 233. 

One of our requirements is that the selection will be unbiased with respect to environment. This means that we should not choose to ignore a region where we have fewer galaxies than IFUs. It also means that we need to overlap tiles in denser regions on the sky to achieve similar completeness in dense environment as in voids. The right panel of Figure \ref{fig:tiling} shows the results of our adaptive tiling routine. In this scheme we use a `gravitational' method to assign multiple overlapping plates to the densest regions. The method starts with an evenly distributed overlapping tiling of $\sim$2000 tiles. The target galaxies then `pull' tiles as an $1/r^2$ law, with the mass of already assigned targets set to zero. The velocity with which the plates are allowed to move is damped to prevent oscillation of the tiles and a mild repulsive force between the plates is included. This has the effect of pulling plates towards the over dense regions and away from the voids. Currently only plates with $>$ 7 targets are included in the final tiling. 

This adaptive tiling scheme is highly effective at increasing our coverage of the densest regions without a significant loss of either IFU allocation efficiency or the ability to assign the correct size IFU to each target. The IFU allocation efficiency is 97.8\% (which increases to 98.5\% when the ancillary targets are included; see \S\ref{sec:anc}) and the completeness within the tiled footprint is 87\% (which decreases to 85\% when the ancillary targets are included), which compares well with the non-overlapping tiling which has a 93\% efficiency and a 60\% completeness. In the adaptive scheme 93\% of the galaxies are allocated IFUs matching the target size and 77\% are allocated IFUs $\ge$ 1.5 or 2.5 \Re~compared to 96.7\% and 78\% for the non-overlapping tiling. Surprisingly we see that the overlapping tiling has an increased allocation efficiency but this is simply the result of the removal of tiles that overlap the edge of the imaging region. It also produces a slight decrease in the allocation of IFUs with the desired size. This decrease is caused by the increased completeness, resulting in a reduction in the average number of available IFUs on each tile, making it harder to find galaxies with target sizes matching the available IFU sizes.  

As well as providing a more even sampling of dense environments our adaptive (overlapping) tiling scheme also enables us to improve observing efficiency. Because we have more total tiles over our area, we have more tiles in the LST regions with fewer overall tiles than in the non-overlapping case, and so can observe the easier tiles in those regions, i.e. ones that are at lower airmass and take less exposure time to complete. Our survey simulations suggest that the adaptive tiling will result in an increase of almost 10\% in the total number of plates that we can complete during the survey.

\subsection{Quality Control and Visual Inspection}
\label{sec:QA}

\begin{figure*}[ht]
 \begin{center}

\includegraphics[width=13cm]{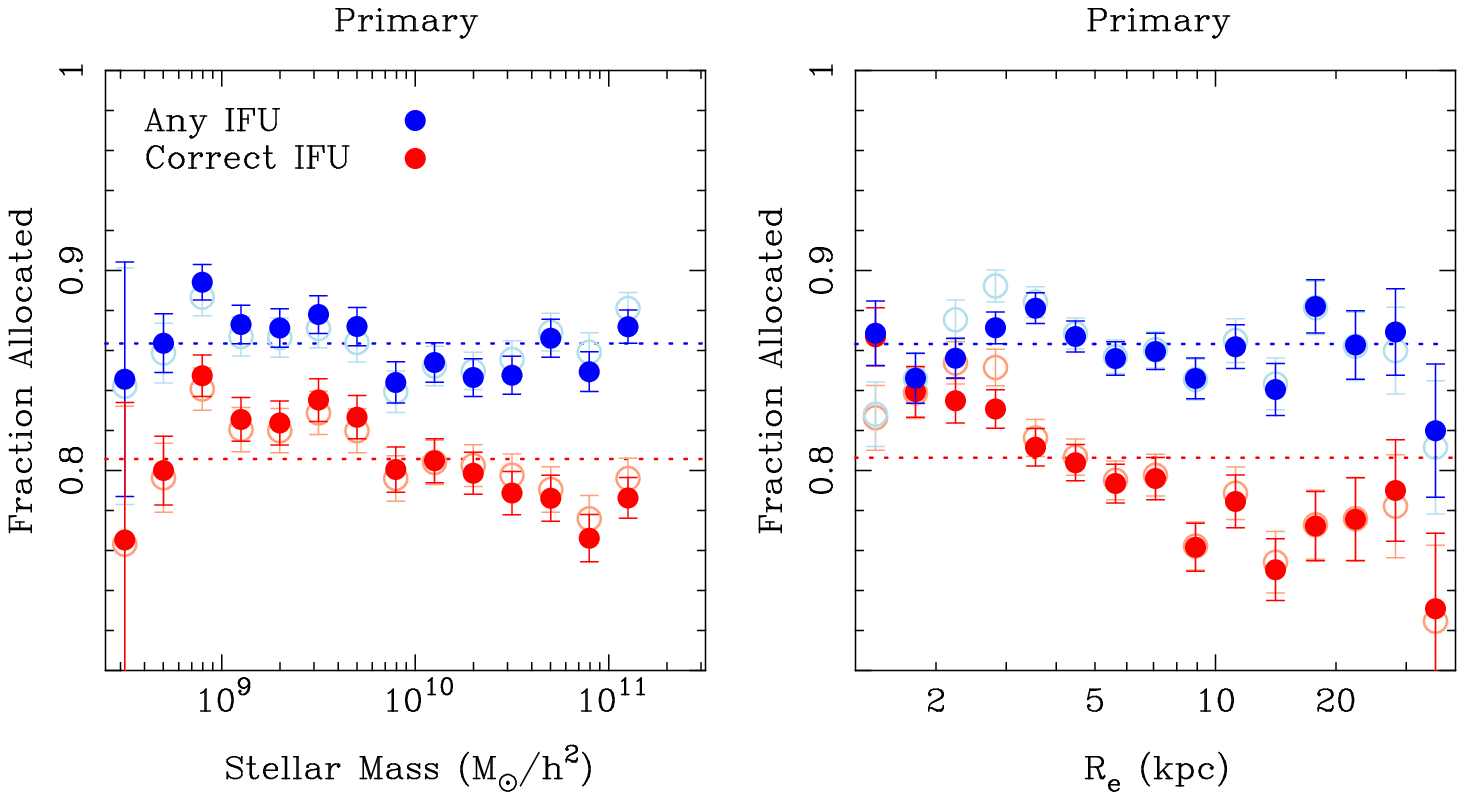}
\\

 \vspace*{0.2cm}

\includegraphics[width=13cm]{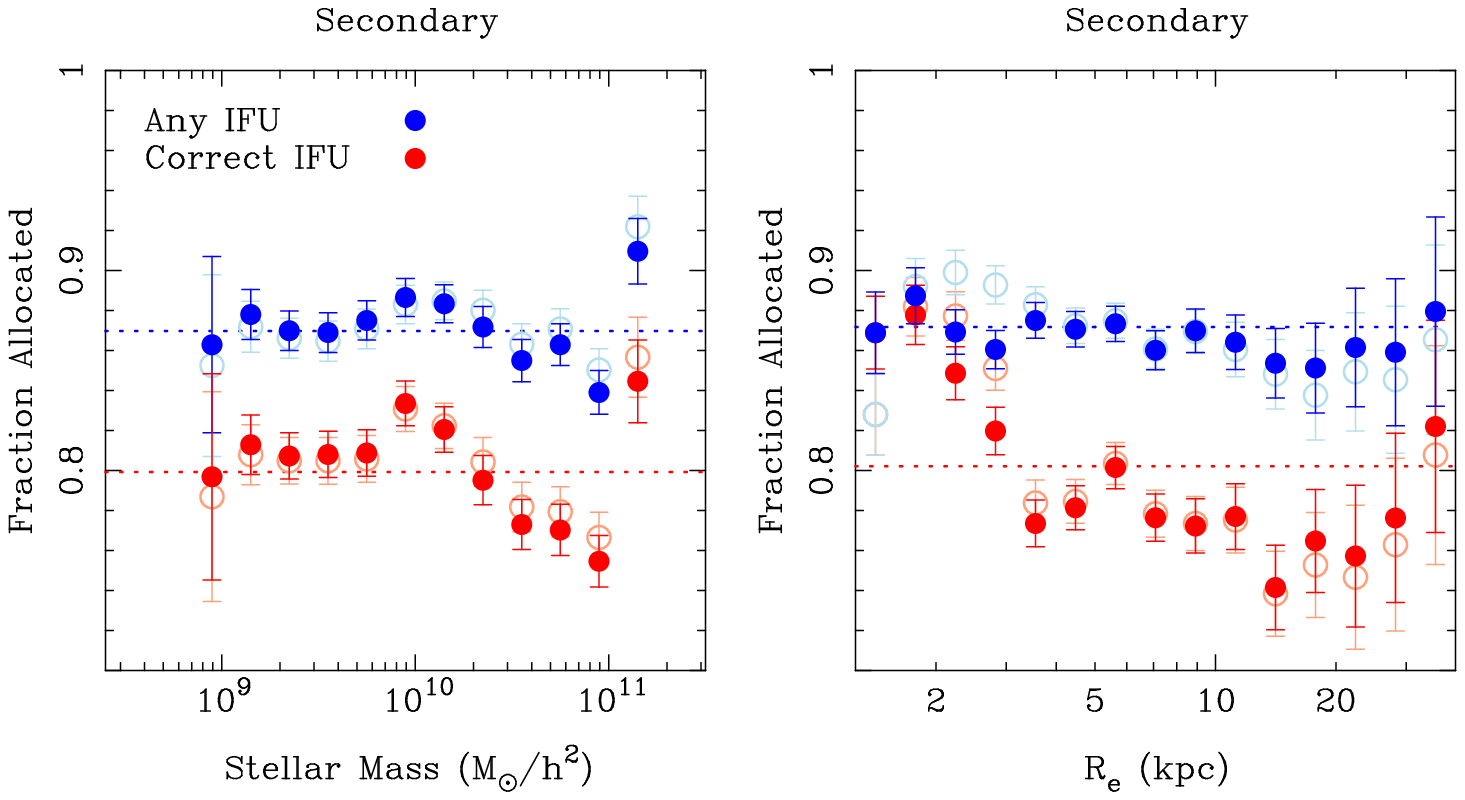}

 \vspace*{0.1cm}
 \caption{\small The fraction of galaxies allocated an IFU (blue) and the fraction allocated an IFU of the correct size (red) as a function of stellar mass (left) and $R_e$ (right) for the Primary (top) and Secondary (bottom) samples. The dashed lines show the fraction irrespective of stellar mass or size. The open lighter colored points show the raw fractions, whereas the filled darker points show the fractions corrected using the IFU size allocation bias weight (see text for details).}
  \label{fig:bunalloc}
 \end{center}
 \end{figure*}

Before allocating IFUs to potential targets we undertook a set of visual inspections of the target catalog to make sure the photometry and hence selection of the targets was reliable. We inspected all galaxies where there was an indication that there may be an issue with the photometry or redshift. Specifically where the Sersic and Elliptical Petrosian \Re\ differed by a factor of 3 or more, the Elliptical Petrosian \Re\ $>$ 20'', the photometric and spectroscopic (i.e. redshift) centers differed by more than 1'', or the redshift source was not the SDSS. This is a total of 4,292 galaxies. For each of these we looked at the imaging, light profiles and spectra for all of these targets. Galaxies were flagged as bad where the photometry had clearly and significantly failed, due to e.g.\ bad imaging, bad deblending with a nearby bright start or galaxy, or a catastrophic background subtraction issue. Galaxies were also flagged where the redshift allocated clearly corresponded to a different galaxy. Finally we provided a new center if the catalog center was obviously incorrect, mainly due to the presence of a foreground star, bright sub-region, or strong dust lanes.

To check that our preselection was catching the vast majority of issues we inspected 500 random targets not already flagged as bad finding no major photometry or centering issues. However, since we do not wish to waste an IFU as the survey proceeds we inspect the targets on each tile before we use that tile to drill any plate that will be observed, again flagging galaxies in the same way. Overall the inspection done to date we have flagged 1227 galaxies as bad and recentered 1189, each amounting to $<$3\% of the total targets. 

Finally we flag the 983 targets lying within 25\arcsec\ of a bright star ($r <$ 14 mag) using the APASS DR8 \citep{Henden12} and Tycho 2 \citep{Hog00} star catalogs. No targets flagged as bad or close to a bright star are eligible for IFU allocation. 

\subsection{IFU Allocation}
\label{sec:allocation}

\begin{figure*}[ht]
 \begin{center}

 \includegraphics[width=13cm]{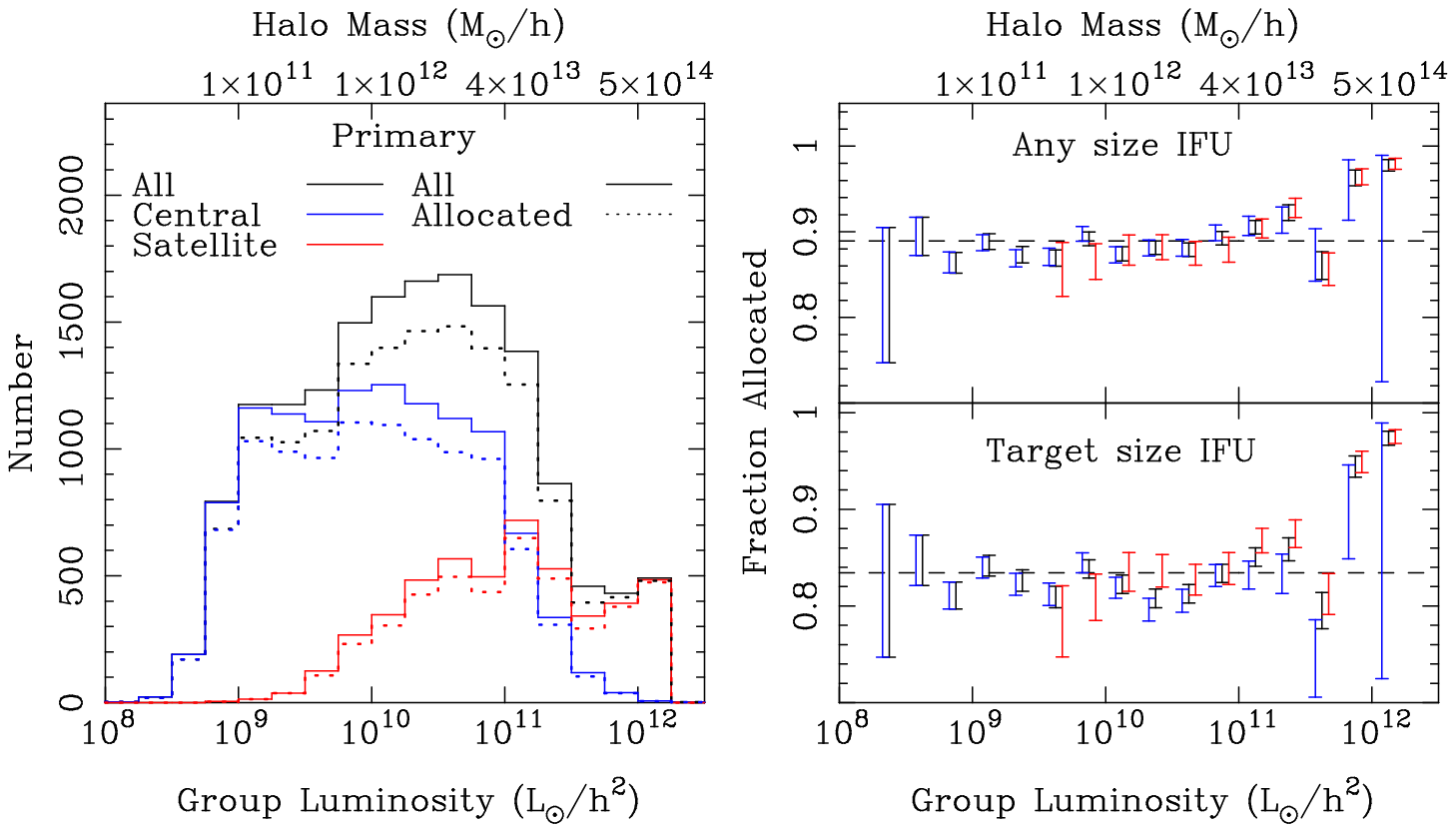}

 \vspace*{0.25cm}

 \includegraphics[width=13cm]{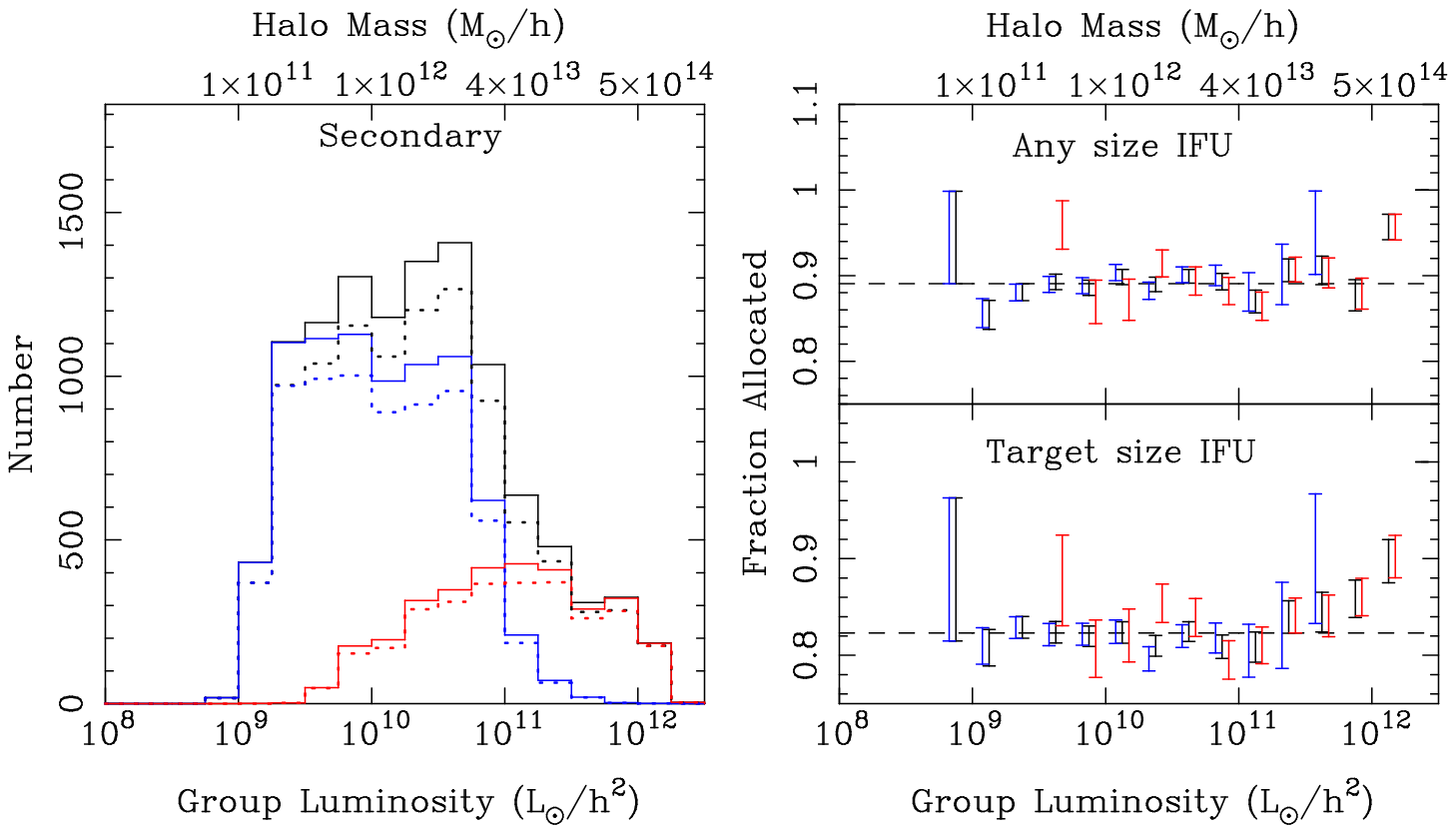}

 \vspace*{0cm}
 \caption{\small Left panels: The distribution in the number of galaxies targeted as a function of group luminosity for the Primary (top) and Secondary (bottom) sample. The solid histograms show all targets and the dotted just those allocated an IFU. We additionally divide the targets into centrals (blue) and satellites (red). Right panels: The fraction of galaxies allocated an IFU of any size (top sub panel) and with a size greater than or equal to the target size (bottom sub panel) as a function of group luminosity for the Primary (top) and Secondary (bottom). All galaxies are shown in black, centrals in blue and satellites in red as before.}
  \label{fig:bunallocenv}
 \end{center}
 \end{figure*}

Once we have broken the survey area into tiles, the next step is allocating individual IFUs to galaxies in each tile. Our method for this procedure is designed to maximize the allocation of IFUs to galaxies of the appropriate size. It proceeds as follows:
\begin{enumerate}
\item All galaxies within a given tile are selected.
\item Galaxies that collide with the center post (within 150\arcsec) are removed and one of every colliding pair (within 120\arcsec) selected at random is removed.
\item For each IFU size, say 19, galaxies that require a 19-fiber IFU to reach their target radius (e.g. 1.5 or 2.5 \Re) are selected. Galaxies with a target radius smaller than the 19-fiber IFU are included in the 19-fiber IFU allocation (galaxies larger than the 127-fiber IFU are likewise assigned to the 127-fiber IFU).
\item All available 19-fiber IFUs are assigned to these galaxies.
\item If there are fewer galaxies that need 19-fiber IFUs than available 19-fiber IFUs, the remaining IFUs are put into a pool to be assigned later.
\item This process is repeated for the remaining IFU sizes.
\item The unassigned IFUs are then allocated to the remaining galaxies closest in target size, beginning with the largest galaxies and the largest remaining IFUs, and then working downwards.  
\end{enumerate}

Galaxies from all the main samples are treated identically (i.e. none are preferred) except those Secondary galaxies that have been removed by down-sampling to get the correct relative number of Primary and Secondary targets. These randomly removed Secondaries will only be allocated an IFU if there are no available targets from another sample.  

This method will produce a sample that has an angular size distribution that matches as closely as possible the IFU size distribution. Since we have optimized the IFU size distribution (see \S\ref{sec:bunsizedist}) to match the actual galaxy target size distribution we expect that any bias introduced will be small. However, we expect some bias to be introduced given that we are using a 2,4,4,2,5 IFU size distribution when 2,4,3,2,5 was slightly more optimal and that we only have a small integer number of each size IFU, which is highly unlikely to match the desired IFU size distribution perfectly. However, any such bias can very easily be corrected by comparing the relative number of IFUs of a given size required for the sample to the number available (see \S\ref{sec:allocweight} for details of these weights). We demonstrate that the bias is indeed small and corrected by appropriate weights in Figure \ref{fig:bunalloc} by determining the fraction of galaxies that are allocated an IFU as a function of stellar mass and size (\Re). 

Figure \ref{fig:bunalloc} shows the results of applying our allocation method to the combined Primary+ and Secondary samples using the adaptive tiling presented in \S\ref{sec:tiling}, and our optimized 2,4,4,2,5 IFU size distribution along with an IFU allocation weight (see \S\ref{sec:allocweight}). In all panels we show the fraction of galaxies allocated an IFU in blue, and the fraction allocated an IFU of the correct size, which is where the IFU is either greater than or equal to the target size (the smaller value between 1.5 or 2.5 \Re\ and the radius of 127-fiber IFU), in red. The raw fractions are shown with the lighter colored open points and the fractions corrected by the IFU allocation weights by the darker filled points. The left panels show these fractions as a function of stellar mass, the right panels as a function of \Re, and the top and bottom panels for Primary and Secondary samples respectively. 

\begin{table*}[ht]
  \begin{center}
    \caption{{\label{tab:allocation}\small IFU allocation results for an adaptive MaNGA tiling of 1800 plates. Numbers in parentheses are fractions.  $R_s$ is 1.5 $R_e$ for the Primary sample and 2.5 $R_e$ for the secondary sample. The target IFU size may still be less than $R_s$ for large galaxies. DS refers to the Secondary sample after it has been down-sampled to it's target density. These results include the allocation of IFUs to Ancillary targets.}}
	\begin{tabular}{ccccc}
\\
	\tableline
    	\multicolumn{1}{c}{Input sample} &
      	\multicolumn{1}{c}{Targets} &
      	\multicolumn{1}{c}{Got IFU} &
      	\multicolumn{1}{c}{R$_{\rm IFU}$ $\geq$ target size} &
      	\multicolumn{1}{c}{R$_{\rm IFU}$ $\geq$ R$_s$} \\
	\tableline

Combined     & 38162  &   29204 (0.765) &  27111 (0.928) & 22413 (0.767) \\
Combined DS  & 32772  &   27806 (0.848) &  25917 (0.932) & 21395 (0.769) \\
Primary+     & 21661  &   18334 (0.846) &  17190 (0.938) & 14286 (0.779) \\
Primary      & 16231  &   13747 (0.847) &  12867 (0.936) & 10514 (0.765) \\
Secondary    & 16533  &   10898 (0.659) &   9947 (0.913) &  8149 (0.748) \\
Secondary DS & 11143  &    9500 (0.853) &   8753 (0.921) &  7131 (0.751) \\
Color-En     &  5430  &    4587 (0.845) &   4323 (0.942) &  3772 (0.822) \\
	\tableline
    \end{tabular}
\end{center}
\end{table*}

We can see that almost 90\% of all galaxies covered by our tiling are allocated an IFU and that there is no trend apparent with stellar mass for either the raw or corrected fractions. In the raw counts there is a weak trend with \Re\, particularly within the Secondary sample, such that small galaxies are slightly more likely to be allocated an IFU than larger ones. This results from the slightly non-optimal match between our IFU size distribution and that required for the final sample, which has been made a little worse by small changes in our sample definition, such as switching from \Mstar\ to \Mi\, adding in the Color-Enhanced supplement and using functional forms to describe our selection cuts, since the 2,4,4,2,5 IFU size distribution was defined. However, the inclusion of the weights completely corrects for this bias leaving no trend with physical size. 

When we turn to looking at the galaxies that are covered by an IFU that is actually larger than their target radius we see the trend with size becomes stronger in both samples. This likely results from the necessity to allocate IFUs to galaxies that don't exactly match the required size when there are not enough galaxies on a tile matching the available IFU sizes. As described above those initially unallocated IFUs are still assigned to the remaining galaxy targets on the tile. Since, for a given mass, intrinsically small galaxies will also have small angular sizes they are easier to cover to their target radius with whatever IFU is available. Whereas an intrinsically large galaxy may only be covered by one of the larger IFUs.  

In Figure \ref{fig:bunallocenv} we show how the allocation efficiency depends on environment. In addition to improving efficiency the main goal of our adaptive tiling scheme is to ensure that the fraction of galaxies allocated IFUs is independent of environment. This is of particular concern in the densest environments where most of the galaxies remain untargeted with just a single plate. We choose to define environment using the Yang et al. SDSS DR7 group catalog \citep{Yang07}, splitting our target galaxies into centrals and satellites and by group luminosity. We also indicate the expected halo mass one might associate with group luminosity using the determinations from the catalog. 

The left panels of Figure \ref{fig:bunallocenv} show the distribution of the numbers of all the selected Primary and Secondary sample galaxies (solid histograms) and those that were allocated an IFU (dotted histograms). Since we have an approximately flat distribution in stellar mass we end up with an approximately flat distribution in halo mass for the central galaxies, with slightly extended tails to high and low halo mass. The extended tail is particularly noticeable at the high halo mass end where the scatter in the halo mass hosting a given stellar mass galaxy is larger. Satellites occupy groups of higher luminosity or mass than centrals with the same stellar mass and so as expected the distribution is offset from the centrals. 

The right hand panels show how the fraction of galaxies allocated an IFU depend on group luminosity, again split by central and satellite designation. For the Primary sample there is very little dependence in the allocated fraction below $10^{11} \Lsun$, but above that there is a general trend of increasing allocation fraction with increasing group luminosity, particularly at the highest luminosities. This trend is caused by  only being able to allocate tiles in integer numbers and the removal of excess tiles with fewer than 7 allocated targets. For galaxies in lower luminosity groups, which are covered on average by $\sim 1.8$ tiles, the removal of a tile with just a few allocated targets has a significant impact on completeness in that region. Galaxies in the most luminous groups are typically in the densest regions and are covered by significantly more tiles (an average of 10.7 for $10^{12} \Lsun$ groups). As a result the  removal of a single tile with fewer then 7 allocated galaxies in such a dense region has a negligible effect on the allocation completeness leaving the fraction of galaxies allocated an IFU close to 100\%. For the Secondary sample we see almost no dependence of the allocation fraction on group luminosity apart from in the most luminous groups ($>10^{12} \Lsun$). The less significant trend simply reflects the smaller number of Secondary sample galaxies and so only the most luminous groups have enough of an effect on the overall sample density to impact the allocation fraction.

Finally, in Table \ref{tab:allocation} we present the numbers of potential targets and allocated IFUs for our final adaptive tiling of 1800 MaNGA tiles that covers the full DR7 area, after excluding tiles containing fewer than seven targets. We have assumed an IFU size distribution of 2,4,4,2,5 and the Primary, Color-Enhanced and Secondary samples as described above. 

Simulations of the survey observing strategy suggest that MaNGA can complete $\sim$600 plates in 6 years, so simply dividing the numbers in this table by 3 will give an indication of the size of the final MaNGA samples and show we should reach our goal of 10,000 galaxies for the three main samples.

Over the full tiling of 1800 tiles we are unable to allocate 453 IFUs due to a lack of targets, which is 1.5\% of the available IFUs. These unallocated IFUs will be allocated to repeated observations of galaxies on overlapping tiles as a test of our data, or finally on the rare occasions this isn't possible, to galaxies that lie just outside our main selection criteria.

\subsection{Ancillary Targets}
\label{sec:anc}

In order to enhance the  broader scientific goals of the MaNGA survey a call for proposals within the collaboration was issued for Ancillary targets that would use a small fraction of the MaNGA IFUs. These programs typically target rarer classes of galaxies that represent important phases in galaxy evolution, but are underrepresented in the three main samples (e.g. luminous AGN, mergers, central galaxies in massive halos) or they target galaxies useful in testing, calibrating, or understanding the main MaNGA data (e.g. high spatial resolution in nearby galaxy, galaxies with calibrated kinematics). The goal for these programs was to make use of allocated IFUs as well as a small fraction of IFUs that could have been allocated to main sample galaxies.

The IFU allocation for the chosen Ancillary programs was made after the allocation to the three main samples. Each program provided a list of targets ranked by priority as well as an IFU size requirement for each target. In addition to the internal priority within each program, each program was prioritized based on how it was ranked by the internal Ancillary TAC. Throughout the ancillary allocation procedure IFUs were allocated in order of these priorities.

The ancillary allocation proceeded in three passes. On the first pass we allocate as many of the unallocated IFUs that match the target size as possible. On the second pass we re-allocate IFUs that have already been allocated to Secondary galaxies that didn't pass the down-sampling cut (identified as RANFLAG = 0 in the catalog) that have the required size. On the third pass we re-allocate IFUs that have been allocated to main sample targets (i.e. Primary+ and Secondary post down-sampling) up to a certain fraction. Since some of the Ancillary targets were already in the main samples and just required a different IFU size the final step is to reallocate the IFUs removed from these targets to a new Primary+ or Secondary sample target of a suitable size if one is available. 

In total we allocate 5.1\% of the available IFUs to ancillaries, 3\% to entirely new ancillary targets and 2.1\% to already allocated targets in the main samples that required a larger IFU size. As a result the fraction of Primary+ and Secondary sample galaxies allocated an IFU reduces from 86.6\% to 84.8\%. The fraction of unallocated IFUs decreases from 2.3\% to 1.5\%. 

Detailed descriptions of the individual ancillary programs are available on the SDSS data release pages from DR13 onwards.

\section{Considerations when using the MaNGA samples}
\label{sec:considerations}

\begin{figure}[h]
 \begin{center}

 \includegraphics[width=7.5cm]{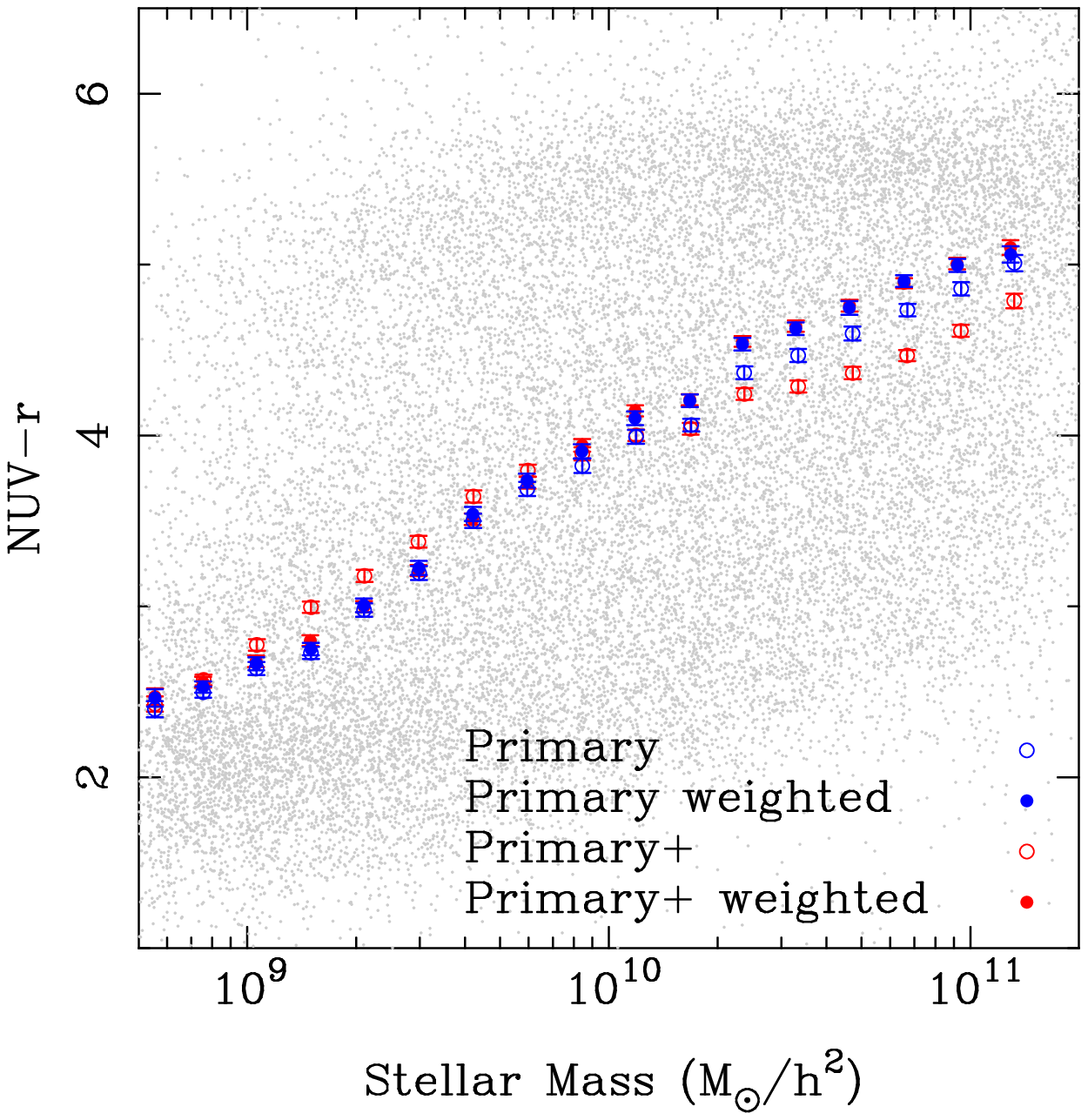}
 \includegraphics[width=7.5cm]{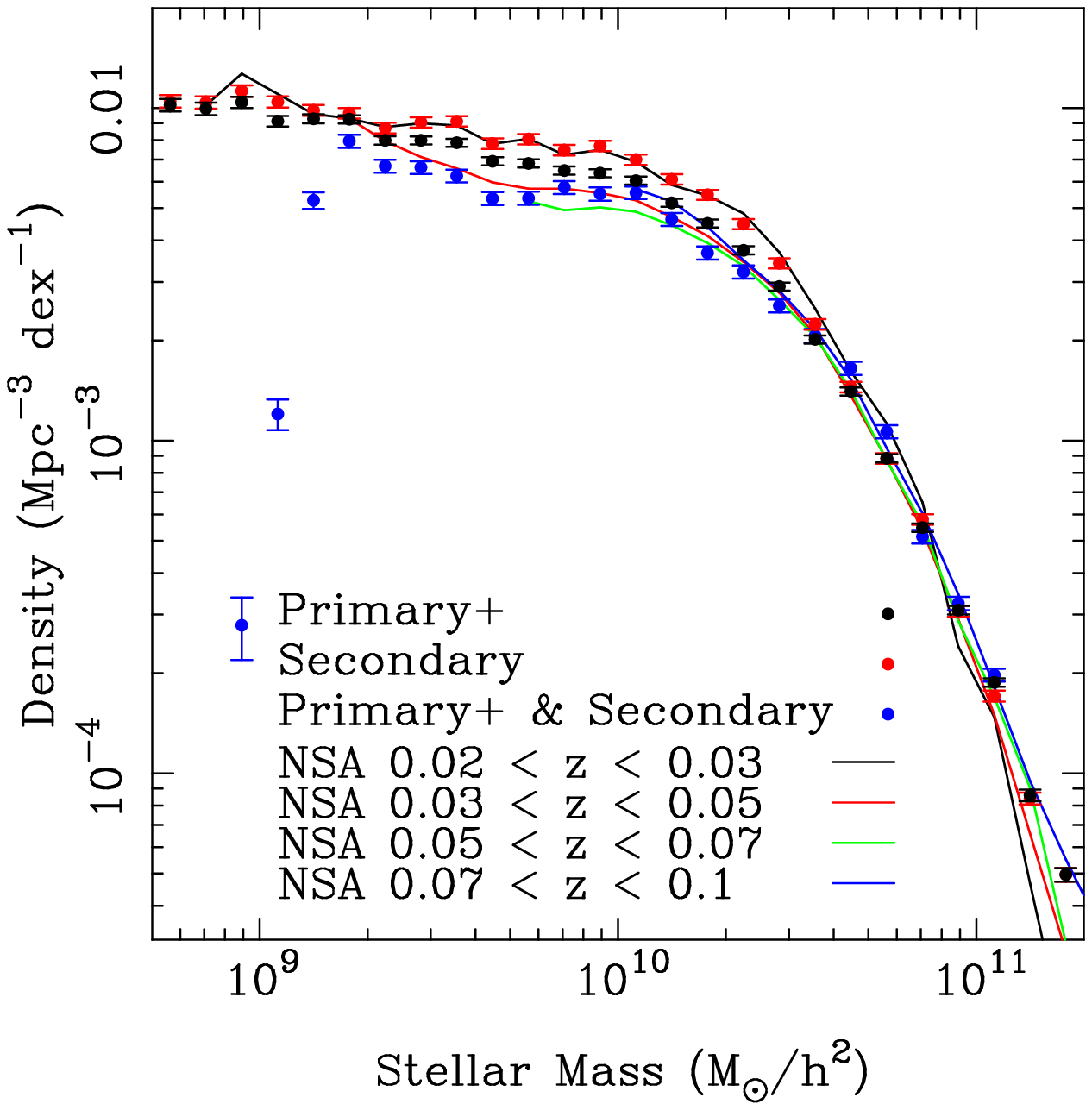}

 \vspace*{0cm}
 \caption{\small Volume weight corrections - Top: The $NUV-r$ colors of the Primary+ galaxies as a function of stellar mass. The open blue and red points show the median colors for the Primary and Primary+ samples respectively. The filled red and blue points show the median colors for the two samples after the Vmax weights have been applied, effectively correcting the median colors back to those of a volume limited sample. Bottom: The stellar mass functions of the Primary+ sample (red points), the Secondary sample (blue points) and their combination (black points) compared to stellar mass functions calculated using the full NSA catalog in a series of redshift ranges (lines).}
  \label{fig:vmax}
 \end{center}
 \end{figure}

The MaNGA sample is selected in a somewhat different manner to typical galaxy samples, which are often either magnitude or volume limited, and so there are certain considerations to take into account when using it for science. In this section we describe how to approach using the sample and some things that should be avoided.

\subsection{Volume Weights}
\label{sec:vmax}

As is the case for many galaxy samples the MaNGA sample is not volume limited, although that is by design rather than observational necessity. We have chosen to flatten the number density distribution as a function of \Mi\ so that we have as many high luminosity galaxies as we have low luminosity galaxies. As already discussed this will produce a fairly even S/N on measured quantities as a function of \Mi\ (or stellar mass) but could (and most likely will) introduce a bias for any analysis as a function of anything other than \Mi\, unless corrections back to a volume limited sample are made. Even in a narrow stellar mass bin the distribution of galaxy mass to light ratio (or star formation rate or metallicty etc) will not be the same as the distribution in that mass bin for a volume limited sample, even if the mass bin were very narrow.

Fortunately the MaNGA selection is designed to make the calculation of the required volume weights very simple since we know for every galaxy the redshift range over which it could have been selected. At a given \Mi\ (or \Mi\ and $NUV-r$ color for the Primary+ sample) the sample is effectively volume limited in that all galaxies with $z_{min}(M_i) < z < z_{max}(M_i)$ are targeted irrespective of their other properties. However, that volume varies with \Mi. Therefore in any analysis of the properties of MaNGA galaxies as a function of anything other than \Mi\ we must correct for this varying selection volume, $V_s(M_i)$ (the volume with $z_{min}(M_i) < z < z_{max}(M_i)$). The simplest approach is just to correct the galaxies back to a volume limited sample by applying a weight (W) to each galaxy in any calculation, such that $W = V_f/V_s$ where $V_f$ is an arbitrary fiducial volume.

For each galaxy in MaNGA we provide ZMIN, ZMAX, SZMIN, SZMAX, EZMIN and EZMAX which are the minimum and maximum redshift each galaxy could have been observed over for the Primary, Secondary and Primary+ samples respectively. So for a given galaxy one can just convert the appropriate (given which sample it's in) $z_{min}$ and $z_{max}$ to the volume ($V_s$) and the weight is 1/Vs.

An additional complication arrises if one is using the Secondary sample on it’s own or in conjunction with the Primary or Primary+ samples. The Secondary sample was down-sampled to the appropriate surface density so that the 2/3 to 1/3 Primary+ to Secondary sample split would be achieved, with an average sampling rate of $\sim$67.1\%. However, we allow unallocated IFUs that cannot be allocated to any other targets to be allocated to Secondaries not included in the down-sampling, which effectively changes the Secondary sampling rate to 76.9\%. These additional galaxies will most likely be on plates with low surface densities of targets and so could be biased towards lower density regions. If you were concerned about such a bias the safest approach is to ignore these additional galaxies and so restrict the Secondary sample to galaxies with RANFLAG = 1. To first order you could then multiply the Secondary weights by 1/0.671 to reflect the down-sampling rate or if you are not concerned about this potential bias then you can use all the Secondaries multiplying the weight by 1/0.769. However, since the down-sampling rate depends weakly on stellar mass (see \S\ref{sec:secprop}) ideally the down-sampling correction to the weight needs to be made on a galaxy by galaxy basis. A full description of how to calculate weights for each sample and combinations of samples is given in Appendix \ref{sec:vmaxfull} and these weights will be made available with DR14.    

Please note that you should never use the Color-Enhanced supplement on its own for statistical population studies. There are regions of color-magnitude space that are not populated in the Color-Enhanced supplement. If you want to include the Color-Enhanced supplement you should use the Primary+ sample (Primary + Color-Enhanced) where all regions of our nominal color-magnitude space are sampled. EZMIN and EZMAX are defined for all galaxies in the Primary+ sample.

Figure \ref{fig:vmax} illustrates the application and importance of the volume weights. In the top panel we plot the $NUV-r$ color of the Primary+ sample as a function of stellar mass along with the mean $NUV-r$ for the Primary and Primary+ samples with and without the volume weights applied. The volume weights used here are calculated relative to a reference volume such that the mean weight is equal to one. Without the volume weights the Primary sample shows bluer mean colors than with the weights, with the difference becoming larger at larger masses. When flattening in \Mi\ as we move from low to high galaxy luminosity we are increasing the selection volume, and hence relative number of galaxies, compared to a volume limited sample. As a result for a given stellar mass you select more blue galaxies, which have a smaller mass-to-light ratio and are hence more luminous, than you would for a volume limited sample. The magnitude of this effect becomes larger at higher masses, corresponding to brighter \Mi\, where the luminosity function is steep and so the volume change required to flatten the number density becomes rapid. Turning to the Primary+ sample we can see a much flatter trend in mean $NUV-r$ than the Primary sample by design, since we've preferentially filled in underrepresented regions of the color-magnitude plane with the Color-Enhanced supplement. When the volume weights are applied the Primary+ mean colors exactly match those of the Primary sample since they are now both weighted back to a volume limited sample.
  
The bottom panel of Figure \ref{fig:vmax} shows the reconstructed stellar mass function of the three main samples (points) compared to mass functions calculated using the full extended NSA sample in different redshift intervals (lines). Here we have used a 1/Vmax weight with Vmax calculated using the area of the survey\footnote{In this figure we have used the full targeting catalog. To make such a plot from observed MaNGA data one must use the appropriate area and completeness corresponding to the set of observed plates.} and the selection redshift limits appropriate for each sample. The NSA mass functions are also calculated using a 1/Vmax weight where Vmax is determined using the stellar mass completeness limit that results from the magnitude limited nature of the NSA. The figure nicely illustrates that the volume weights modify the approximately flat stellar mass density distributions seen in Figures \ref{fig:finalpriprop}, \ref{fig:finalsecprop} and \ref{fig:primplus} to the expected Schechter function shape of the mass function. We can also see good agreement between the mass functions of the individual samples and those calculated using the full NSA in the redshift ranges that overlap well with a given sample. 

Finally Figure \ref{fig:vmax} demonstrates the advantage of being able to select a sample with a constant number density as a function of \Mi.\ Whereas the errors on the NSA mass functions depend strongly on mass (there would be an even stronger dependence with a volume limited sample rather than the magnitude limited NSA) they are essentially constant for the MaNGA sample.

\subsection{IFU allocation weights}
\label{sec:allocweight}

\begin{figure}
 \begin{center}

 \includegraphics[width=7cm]{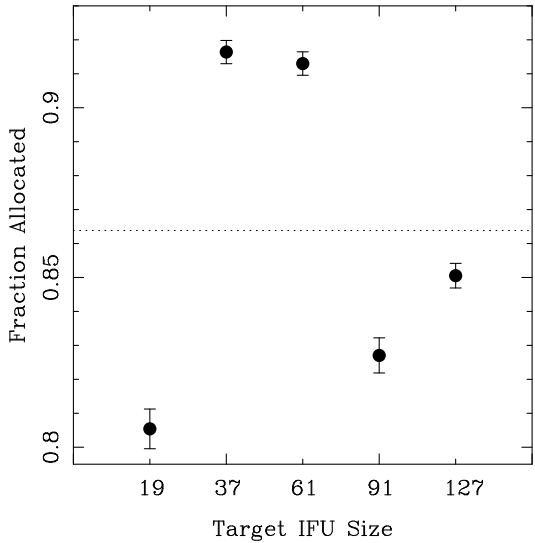}

 \vspace*{0cm}
 \caption{\small The fraction of galaxies allocated an IFU as a function of target IFU size for all galaxies in the Primary+ and Secondary (after down-sampling) samples. The overall fraction is shown as the dotted line. There is a small variation in the allocation fraction with target IFU size as a result of the IFU size distribution not being perfectly optimal.}
  \label{fig:fracifualloc}
 \end{center}
 \end{figure}

To most efficiently cover our targets to the desired radii we select galaxies that match in size to the available IFU sizes on each tile. For several reasons (see \ref{sec:bunsizedist} and \ref{sec:allocation} for a discussion of these), and despite careful optimization, the MaNGA IFU size distribution does not exactly match the distribution required for our final samples. This slight mismatch is demonstrated in Figure \ref{fig:fracifualloc} where we show the fraction of IFUs allocated as a function of the target IFU size for each galaxy. The allocation fraction varies from almost 0.85 to 0.95 with targets requiring the 19 fiber IFU being the least likely to be allocated an IFU and targets requiring the 37 fiber IFU the most likely. Although the variation is small, less than 5\% from the mean, it will introduce a slight angular size selection bias, which translates to a physical size bias (see Figure \ref{fig:bunalloc}) and hence a surface mass density bias etc. 

\begin{table}[h]
  \begin{center}
    \caption{\label{tab:allocweight}}
	\begin{tabular}{cc}
\\
	\tableline
      	\multicolumn{1}{c}{IFU Target Size} &
      	\multicolumn{1}{c}{Allocation Weight} \\
	\tableline
        19 & 1.075\\
        37 & 0.945\\
        61 & 0.949\\
        91 & 1.047\\
        127 & 1.018\\
	\tableline
    \end{tabular}
\end{center}
\end{table}

However, once again it is very simple to correct this bias with a weight. We define the allocation weight for a galaxy with a given IFU target size $S_{IFU}$ as $W_a(S_{IFU}) = A(S_{IFU})/A_{all}$ where $A(S_{IFU})$ is the allocation fraction for that IFU target size and $A_{All}$ is the allocation fraction for all targets, i.e. the ratio of the points to the line in Figure \ref{fig:fracifualloc}. Table \ref{tab:allocweight} gives the weights for each IFU target size for a combined sample of the Primary+ and Secondary (after down-sampling, RANFLAG = 1) targets. We combine the samples since the IFU allocation procedure is blind to the them and so the correction should be the same for them all. We don't include in this calculation the Secondary targets that are initially excluded by down-sampling since they are allocated only when there are no other targets remaining on a tile and so will have a different size bias, as well as the likely environment bias we discussed in \S\ref{sec:vmax}.

Figure \ref{fig:bunalloc} shows that applying these weights removes any bias in the allocation fraction with \Re.

\begin{figure}[h]
 \begin{center}

 \includegraphics[width=7cm]{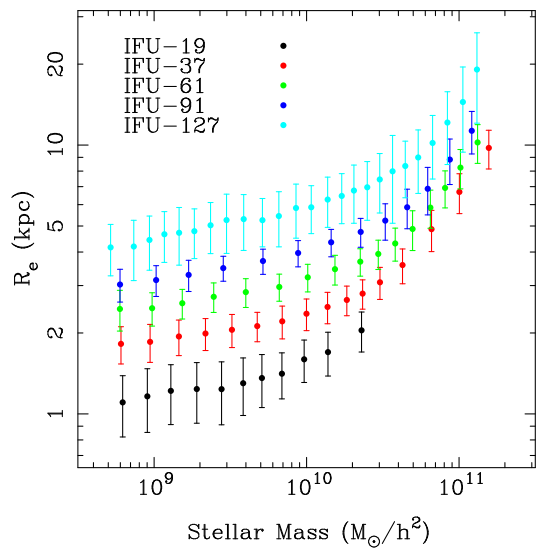}

 \vspace*{0cm}
 \caption{\small The mean \Re\ as function of stellar mass for Primary sample galaxies split by IFU target size. Larger IFUs are typically assigned to intrinsically larger galaxies of a given mass. This could introduce a significant bias if only a subset of the IFUs were used for a given analysis.}
  \label{fig:reIFU}
 \end{center}
 \end{figure}

 \begin{figure*}[ht]
 \begin{center}

 \includegraphics[width=18cm]{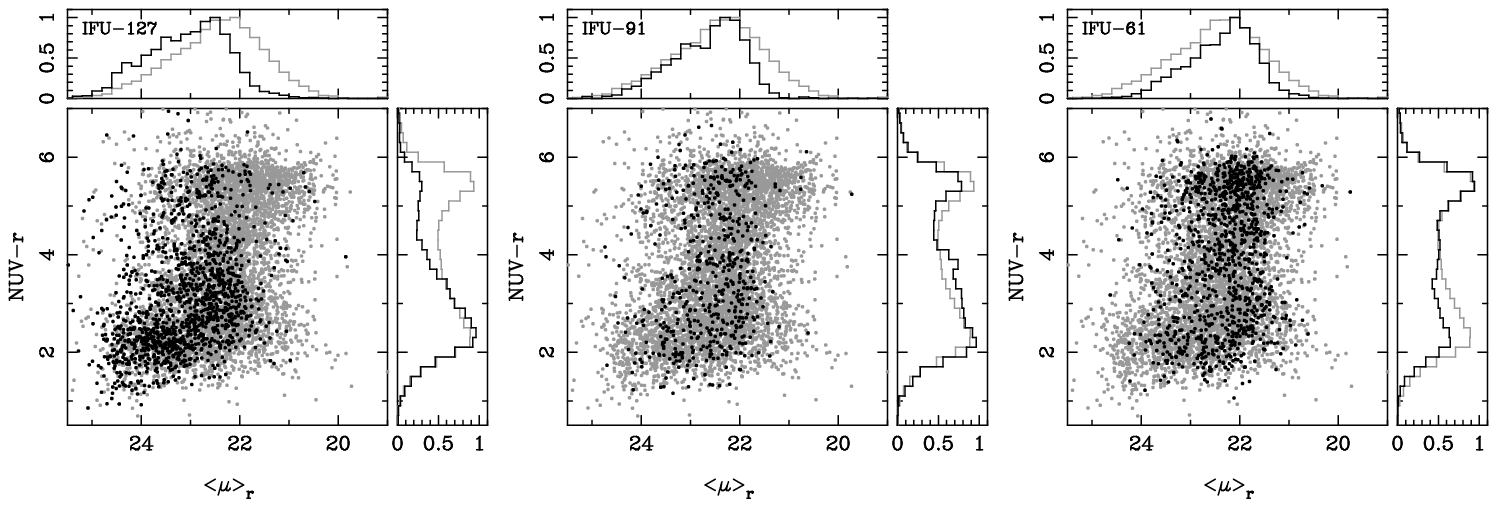}
\\
 \includegraphics[width=12cm]{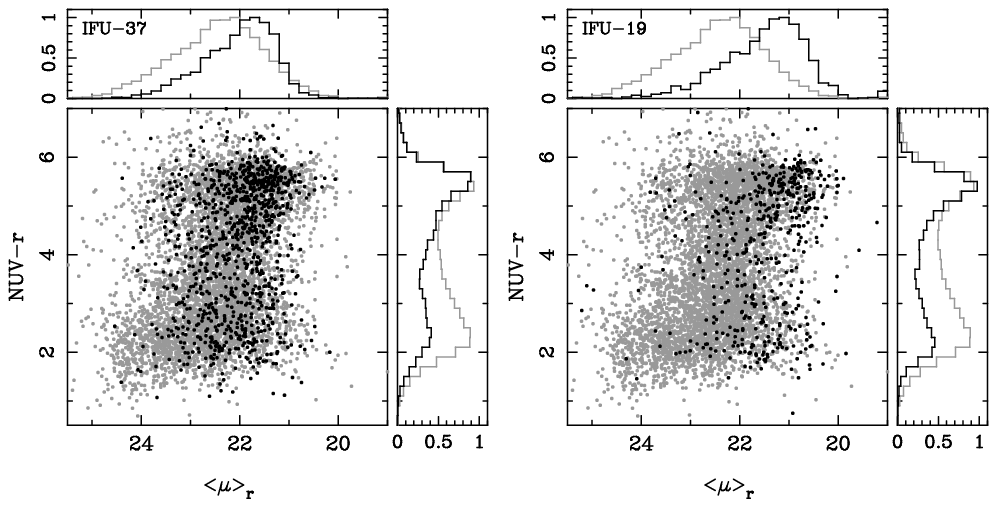}

 \vspace*{0cm}
 \caption{\small The distribution of Primary sample galaxies in the $NUV-r$ vs r-band surface brightness plane. The full sample is shown as the light grey points and histograms in every panel, whereas the black points and histograms in each panel are for a different IFU size. The large IFUs are more likely to be allocated to low surface brightness blue galaxies, whereas the small IFUs are more likely to be allocated to high surface brightness red galaxies.}
  \label{fig:colsbIFU}
 \end{center}
 \end{figure*}
   
\subsection{Selecting by IFU size - don't do it!}

In many cases the precision with which one could make a certain measurement from the MaNGA data will depend on the number of spatial resolution elements there are covering a galaxy. Since MaNGA uses a range of IFU sizes the number of spatial resolution elements varies by a factor of 6.7 between the largest and smallest MaNGA IFUs. It may well then be tempting to use only the largest IFUs for a given science analysis. However, this will introduce a significant bias in the sample selected. For a given \Mi\ (or mass) the redshift range over which the MaNGA targets are selected is narrow corresponding to a small range in spatial resolution being probed. This means that larger IFUs are typically being assigned to intrinsically larger galaxies.

Figures \ref{fig:reIFU} and \ref{fig:colsbIFU} illustrate the correlation between and IFU size and intrinsic galaxy size within MaNGA and the consequences of considering only certain IFU sizes. Figure \ref{fig:reIFU} shows the mean \Re\ as a function of stellar mass for Primary sample galaxies split by IFU target size. As expected at every mass the galaxies requiring a 19 fiber IFU are on average significantly smaller than those requiring a 127 fiber IFU by up to a factor of 4. The consequences of this are shown in Figure \ref{fig:colsbIFU} where we show the distribution of the Primary galaxies in the $NUV-r$ color versus $r$-band surface brightness plane for galaxies targeted with each size of IFU. The largest IFUs are biased towards low surface brightness and hence towards bluer galaxies, whereas the opposite is true for the smallest IFUs  with a continuous trend between the two extremes for the intermediate sized IFUs. 

As a result we would advise that great caution is taken to ensure that the aims of any analysis that uses only a subset of the IFU sizes is independent of such biases.

\subsection{Selecting by Inclination}

Selecting subsamples by inclination will likely be desirable for many science analyses using MaNGA data, whether it is favoring or avoiding edge- or face-on galaxies. Such subsamples introduce no bias to the distribution of the properties of the subsamples as long as the standard volume weights are used. One might worry that the increase in extinction as galaxies become viewed more edge-on could introduce a bias and indeed without the weights there is a small inclination bias in the sample, such that there are fewer edge-on galaxies than face-on galaxies at a given stellar mass than one would find in a volume limited sample. This bias results from the extra extinction of an edge on galaxy results in a fainter \Mi\ at a given \Mstar\ meaning it would only be selected over a lower and narrower redshift range and hence a smaller volume. The same would happen if we selected by extinction specifically, or by color, or SFR, and is simply the result of using \Mi\ in the selection rather than mass. All of these biases are removed by using the volume weights. The exception to this comes at low masses, which we discuss next.  

\subsection{The lowest mass galaxies}

The only circumstances where the volume weights may not completely remove selection bias as intended is at the lowest masses. Since our selection is based on \Mi\ (or \Mi\ and color), at a fixed stellar mass some fraction of the galaxies may not be in our sample at all. For instance red galaxies, either because they are old, dusty or inclined. This is simply the familiar stellar mass incompleteness present in all samples selected using luminosity. The color dependent \Mi\ that we use at faint magnitudes (\S\ \ref{sec:selfn}) greatly reduces the magnitude of this effect but it will to be an issue for the reddest galaxies. We estimate that we are virtually complete for \Mstar\ $> 5 \times 10^8 \Msun$ for the Primary and Primary+ samples and \Mstar\ $> 2 \times 10^9 \Msun$ for the Secondary sample. We note that we are most likely still incomplete for the most inclined galaxies at these stellar mass limits, where the $i$-band extinction may be higher than for face on galaxies by close to 1 magnitude. As such, analyses making use of only the most inclined systems should approach these completeness limits with caution.     

\subsection{Fiber/IFU collisions}

 As already discussed, the overall fraction of target galaxies allocated an IFU is $\sim$84\% with only a very weak dependence on size, stellar mass or halo mass. However, as a result of collisions between IFUs in MaNGA and between spectroscopic fibers in the original SDSS I/II spectroscopy \citep{Blanton03} there is a significant decrease in completeness for pairs of galaxies with small separations $<$120\arcsec\ (8$h^{-1}$ kpc to 115$h^{-1}$ kpc over the MaNGA redshift range). The fraction of MaNGA targets that are in a collision (pairs with separations $<$120\arcsec) is 8\% and for these targets the fraction allocated an IFU is decreased to 71\%. 

In addition there is a further incompleteness within the MaNGA target catalog as a result of spectroscopic incompleteness in the extended NSA. This incompleteness is mainly the result of fiber collisions within the original SDSS spectroscopy, which occur at pair separations $<$55\arcsec, and results in an overall spectroscopic completeness of the extended NSA of $\sim$98\%. We estimate that for MaNGA targets with pairs separated by $<55$\arcsec\ the spectroscopic completeness of the target catalog is reduced at most to 69\%. This is a lower limit since to make this calculation we have assumed that galaxies with missing redshifts have the same redshift as their nearest neighbor. If that were not the case the likelihood of both galaxies in a close pair being targeted by MaNGA is reduced. The final completeness within 55\arcsec\ will then be further reduced as a result of the the MaNGA IFU collisions to $\sim$47\%.

While these reductions in completeness for close pairs may seem large, it is important to remember that this is affecting a very small fraction of the full galaxy population and so for most purposes making global corrections for incompleteness (one for the extended NSA and one for MaNGA IFU allocation) should be sufficient. However, if one is studying close pairs and the study requires an absolute normalization, such as would be needed for merger fractions as a function of separation or very small scale clustering, further corrections should be made \citep[e.g.][]{Patton16,Guo12}.

\section{Summary and Conclusions}

Given the bounds of the SDSS telescope and the BOSS spectrograph the MaNGA sample has been designed alongside the IFU size distribution and allocation strategy to efficiently produce a sample of approximately 10,000 galaxies with a relatively even sampling of color-magnitude space and galaxy environment, along with radial coverage to either 1.5 or 2.5 \Re. This selection has been achieved using simple criteria based purely on $i$-band absolute magnitude (and $NUV-i$ color for a small subset of targets) and redshift. 

We have found that to most efficiently target such a sample requires the following:

\begin{enumerate}
\setcounter{enumi}{0}
\item A range in IFU sizes, to account for the intrinsic variation of galaxy size even at fixed mass and redshift. 

\item The selection of more luminous or more massive galaxies at higher redshift, to produce an angular size distribution that is largely independent of mass. 

\item The density of targets, which is determined by the volume probed at any given \Mi\, must be sufficiently large so as to provide enough targets as to be able to efficiently allocate IFUs of the correct size thus minimizing unused IFUs, and not too large so as to produce too wide a range in galaxy angular size.

\item Within those constraints the targets should be at as low a redshift as possible to maximize S/N and spatial resolution.
\end{enumerate}

Following our optimization we select an IFU size distribution with 2x19-fiber, 4x37-fiber, 4x61-fiber, 2x91-fiber and 5x127-fiber IFUs and design two samples, the Primary+ sample with coverage to 1.5 \Re and the Secondary sample with coverage to 2.5 \Re. This dual sample approach was designed to allow exploration of the dark matter dominated regime and poorly studied outskirts of galaxies, while still also producing a sample that is well spatially sampled in the high S/N inner regions of galaxies that contain the bulk of the stellar mass.

The Primary+ sample, making up two thirds of the targets, is made up of the Primary sample selected purely by \Mi\ and redshift and the Color-Enhanced supplement, which includes an additional selection in $NUV-i$ color and fills in low-density regions of color-magnitude space. The median r-band per fiber S/N at 1.5 \Re\ for the Primary sample is expected to be 8.3 with little dependence on galaxy mass other than for the lowest mass galaxies. The median spatial resolution is 0.35 \Re\ and 1.37 kpc (1.2 kpc for \Mstar\ $< 10^{10} \Msun$, 3.8 kpc for \Mstar\ $> 10^{11} \Msun$).  

The Secondary sample, making up one third of the targets, is, like the Primary sample, selected purely on \Mi\ and redshift but targets higher redshift galaxies in-order-to provide coverage to 2.5 \Re\ with the same size IFUs. As a result both the spatial resolution and S/N is reduced. The median resolution is a factor of 1.7 poorer, and the S/N is just 2.3 per fiber at 2.5 \Re. For this sample the primary science is expected to come from emission line studies, where the line S/N will be much higher, or through stacking multiple fibers in single galaxies or from samples of galaxies.

While the even sampling in stellar mass and color greatly enhances the usefulness of the MaNGA sample for studies over the full range of galaxy properties it does require some care to avoid subtle biases being introduced into any population analysis. However, since the selection is simply defined by a redshift range at a given \Mi\ (or \Mi\ and $NUV-i$ color) it is straightforward to calculate a volume weight for each galaxy, that corrects back to a volume limited sample thus eliminating these biases.

It is clear from the MaNGA data that has already been taken that this sample will achieve its objectives of enabling a wide range of statistical analyses of the spatial distribution of galaxy properties in the local Universe, thus providing great insight into the physics of galaxy formation.

The full MaNGA target catalog including many parameters from the NSA and IFU allocation information is publicly available as part of the SDSS data releases from DR13, with the volume weights included from DR14 onwards.
 
\acknowledgments

AW acknowledges support of a Leverhulme Trust Early Career Fellowship.

This research was made possible through the use of the AAVSO Photometric All-Sky Survey (APASS), funded by the Robert Martin Ayers Sciences Fund.

Funding for the Sloan Digital Sky Survey IV has been provided by the
Alfred P. Sloan Foundation, the U.S. Department of Energy Office of
Science, and the Participating Institutions. SDSS acknowledges
support and resources from the Center for High-Performance Computing at
the University of Utah. The SDSS web site is www.sdss.org.

SDSS is managed by the Astrophysical Research Consortium for the Participating Institutions of the SDSS Collaboration including the Brazilian Participation Group, the Carnegie Institution for Science, Carnegie Mellon University, the Chilean Participation Group, the French Participation Group, Harvard-Smithsonian Center for Astrophysics, Instituto de Astrofísica de Canarias, The Johns Hopkins University, Kavli Institute for the Physics and Mathematics of the Universe (IPMU) / University of Tokyo, Lawrence Berkeley National Laboratory, Leibniz Institut für Astrophysik Potsdam (AIP), Max-Planck-Institut für Astronomie (MPIA Heidelberg), Max-Planck-Institut für Astrophysik (MPA Garching), Max-Planck-Institut für Extraterrestrische Physik (MPE), National Astronomical Observatories of China, New Mexico State University, New York University, University of Notre Dame, Observatório Nacional / MCTI, The Ohio State University, Pennsylvania State University, Shanghai Astronomical Observatory, United Kingdom Participation Group, Universidad Nacional Autónoma de México, University of Arizona, University of Colorado Boulder, University of Oxford, University of Portsmouth, University of Utah, University of Virginia, University of Washington, University of Wisconsin, Vanderbilt University, and Yale University.





\appendix
\section{Volume Weight Calculation}
\label{sec:vmaxfull} 

We define a volume for each galaxy in each of the three samples as the co-moving volume enclosed by the redshift selection limits for that galaxy and the tiled area of the survey (7362 deg$^2$), $V_P$, $V_S$ and $V_{P_+}$ for the Primary, Secondary and Primary+ sample galaxies respectively. We also define a fiducial volume $V_F = 1\times 10^6 Mpc^3$, which is approximately the average volume of the MaNGA samples, although this choice is arbitrary.

To correct for the allocation incompleteness we calculate the probability $P$ that a given galaxy is allocated an IFU. For the Primary and Primary+ samples this is simply the ratio of the number of targets allocated an IFU to the total number that could have been, i.e. that lie on a tile\footnote{The completeness will vary as the survey progresses with more and more overlapping tiles being observed with time. As a result, if absolute rather than relative volumes weights are required this probability and hence the weights should be recalculated for each data release.}, and is the same for all galaxies in that sample. So

\begin{equation}
 P_P = N_{Pa}/N_P 
\end{equation}
and
\begin{equation}
 P_{P_+} = N_{P+a}/N_{P+}
\end{equation}

where N is the number of galaxies and subscript P and P+ indicate the Primary and Primary+ samples and subscript a only those allocated an IFU.

For the Secondary sample we must also take into account the effect of the stellar mass dependent down-sampling and so the allocation probability is uniquely defined for each galaxy. Additionally, since in the initial main allocation stage we allocate only to those galaxies that {\it pass} the down-sampling (RANFLAG = 1) and then allocate any remaining unallocated IFUs to those that do not {\it pass} (RANFLAG = 0), we must treat the down-sampled and non-down-sampled Secondaries separately, calculating $P_{Sd}$ and $P_{Snd}$ respectively. So we have

\begin{equation}
 P_{Sd}(M_*) = P_{d}(M_*) \times N_{Sda}/N_{Sd}
\end{equation}
and
\begin{equation}
 P_{Snd}(M_*) = [1-P_{d}(M_*)] \times N_{Snda}/N_{Sd}
\end{equation}

where $P_{d}(M_*)$ is the probability that a given galaxy is included after the stellar mass dependent down-sampling. The allocation probability for the full Secondary sample is then simply
\begin{equation}
 P_{S}(M_*) = P_{Sd}(M_*) + P_{Snd}(M_*).
\end{equation}

 We then combine these allocation probabilities with the maximum volumes to produce the weights for each individual sample as 

\begin{eqnarray}
 &W_P &= V_F / (V_P P_P) \\
 &W_S &= V_F / (V_S P_S) \\
 &W_{Sd} &= V_F / (V_S P_{Sd}) \\
 &W_{P_+} &= V_F / (V_{P_+} P_{P_+})
\end{eqnarray}
and for combinations of samples as

\begin{eqnarray}
\nonumber &W_{PSd} &= V_F / (V_P P_P + V_S P_{Sd}) \\
 &W_{P_+Sd} &= V_F / (V_{P_+m} P_{P_+} + V_S P_{Sd}) \\
 &W_{PS} &= V_F / (V_P P_P + V_S P_{S}) \\
 &W_{P_+S} &= V_F / (V_{P+m} P_{P_+} + V_S P_{S}).
\end{eqnarray}
Where we are combining the Primary+ and Secondary sample we modify the Primary+ volume ($V_{P_+m}$) so we do not double count in the rare instances where the upper Primary+ redshift limit overlaps with the lower Secondary redshfit limit.

Applying these weights to MaNGA targets allocated an IFU within the tiled area of the full MaNGA target catalog will produce a volume limited sample with a volume of $1 \times 10^6 Mpc^3$, and so densities are simply $N/1 \times 10^6 Mpc^{-3}$ where $N$ is the weighted number of galaxies. If only a subset of the galaxies are being used, as will most likely be the case, the volume must be scaled by the ratio of the observed area to the fully tiled area of 7362 deg$^2$.




\bibliography{manga_references}

\clearpage

\end{document}